\documentclass[preprint]{elsarticle}
\usepackage[utf8]{inputenc}
\usepackage{moreverb}
\usepackage{times}
\usepackage{amssymb}
\usepackage{amsmath}
\newtheorem{theorem}{Theorem}
\usepackage{natbib}

\bibpunct[,]{(}{)}{;}{a}{,}{,}
\title
      {Calculating risk in functional programming 
}

\author[uminho]{Daniel Murta\fnref{fn1}} 
\ead{pg23205@alunos.uminho.pt}

\author[uminho]{Jos\'{e} Nuno Oliveira\corref{cor1}}
\ead{jno@di.uminho.pt}

\cortext[cor1]{Corresponding author.}
\fntext[fn1]{Partially supported by
	\emph{Funda\c{c}\~ao para a Ci\^encia e a Tecnologia},
	Portugal, under grant number
	\\
	\tt BI1-2012\_PTDC/EIA-CCO/122240/2010\_UMINHO.}

\address[uminho]{HASLAB - High Assurance Software Laboratory\\ INESC TEC / Univ.\ Minho, Braga, Portugal}

\usepackage{url}
\def\FST{{\tiny\begin{bmatrix}
1	&1	&1	&0	&0	&0\\
0	&0	&0	&1	&1	&1
\end{bmatrix}}
}
\def\SND{{\tiny\begin{bmatrix}
1	&0	&0	&1	&0	&0 \\
0	&1	&0	&0	&1	&0 \\
0	&0	&1	&0	&0	&1
\end{bmatrix}}
}
\def\F{{\tiny\begin{bmatrix}
0.5	&0.3	&0	&0.75 \\
0.5	&0.7	&1	&0.25
\end{bmatrix}}
}
\def\G{{\tiny\begin{bmatrix}
0.3	&0.4	&0.1	&0 \\
0.7	&0.2	&0.2	&1 \\
0	&0.4	&0.7	&0
\end{bmatrix}}
}
\def\K{{\tiny\begin{bmatrix}
0.15	&0.12	&0	&0 \\
0.35	&0.06	&0	&0.75 \\
0	&0.12	&0	&0 \\
0.15	&0.28	&0.1	&0 \\
0.35	&0.14	&0.2	&0.25 \\
0	&0.28	&0.7	&0
\end{bmatrix}}
}
\newenvironment{lcbr}{\left\{\begin{array}{l}}{\end{array}\right.}
\def\matlab{\textsc{Matlab}}
\def\kr{\mathbin{\hbox{\tiny${}^\vartriangle$}}}
\def\mean#1{\mathopen{[\![}#1\mathclose{]\!]}}
\def\kcomp{\mathbin{\bullet}}
\def\succ{\mathsf{succ}}
\def\inN{\mathsf{in}}
\def\N{\mathbb{N}\index{Números naturais ($I\!\!N$)}}
\def\msplit#1#2{\left[\frac{#1}{#2}\right]} 
\let\iso=\cong
\let\kons=\underline
\def\mblock#1#2#3#4{\left[\begin{array}{c|c}#1&#2\\\hline#3&#4\end{array}\right]}
\def\eqnnewpage{\end{eqnarray*}\newpage~\vskip-2em\begin{eqnarray*}&& }
\def\fun#1{{\sf #1}}
\usepackage[all]{xy}
\def\larrow#1#2#3{\xymatrix{ #3 & #1 \ar[l]_-{#2} }}
\def\rarrow#1#2#3{\xymatrix{ #1 \ar[r]^-{#2} & #3 }}
\def\lharpoon#1#2#3{\xymatrix{ #3 & #1 \ar@{_{>}}[l]_-{#2} }}

\makeatletter
\@ifundefined{lhs2tex.lhs2tex.sty.read}%
  {\@namedef{lhs2tex.lhs2tex.sty.read}{}%
   \newcommand\SkipToFmtEnd{}%
   \newcommand\EndFmtInput{}%
   \long\def\SkipToFmtEnd#1\EndFmtInput{}%
  }\SkipToFmtEnd

\newcommand\ReadOnlyOnce[1]{\@ifundefined{#1}{\@namedef{#1}{}}\SkipToFmtEnd}
\usepackage{amstext}
\usepackage{amssymb}
\usepackage{stmaryrd}
\DeclareFontFamily{OT1}{cmtex}{}
\DeclareFontShape{OT1}{cmtex}{m}{n}
  {<5><6><7><8>cmtex8
   <9>cmtex9
   <10><10.95><12><14.4><17.28><20.74><24.88>cmtex10}{}
\DeclareFontShape{OT1}{cmtex}{m}{it}
  {<-> ssub * cmtt/m/it}{}

\DeclareFontShape{OT1}{cmtt}{bx}{n}
  {<5><6><7><8>cmtt8
   <9>cmbtt9
   <10><10.95><12><14.4><17.28><20.74><24.88>cmbtt10}{}
\DeclareFontShape{OT1}{cmtex}{bx}{n}
  {<-> ssub * cmtt/bx/n}{}

\newcommand{\Conid}[1]{\mathit{#1}}
\newcommand{\Varid}[1]{\mathit{#1}}
\newcommand{\anonymous}{\kern0.06em \vbox{\hrule\@width.5em}}

\renewcommand{\leq}{\leqslant}

\usepackage{polytable}

\@ifundefined{mathindent}%
  {\newdimen\mathindent\mathindent\leftmargini}%
  {}%

\def\resethooks{%
  \global\let\SaveRestoreHook\empty
  \global\let\ColumnHook\empty}
\newcommand*{\savecolumns}[1][default]%
  {\g@addto@macro\SaveRestoreHook{\savecolumns[#1]}}
\newcommand*{\restorecolumns}[1][default]%
  {\g@addto@macro\SaveRestoreHook{\restorecolumns[#1]}}
\newcommand*{\aligncolumn}[2]%
  {\g@addto@macro\ColumnHook{\column{#1}{#2}}}

\resethooks

\newcommand{\onelinecommentchars}{\quad-{}- }
\newcommand{\commentbeginchars}{\enskip\{-}
\newcommand{\commentendchars}{-\}\enskip}

\newcommand{\visiblecomments}{%
  \let\onelinecomment=\onelinecommentchars
  \let\commentbegin=\commentbeginchars
  \let\commentend=\commentendchars}

\newcommand{\invisiblecomments}{%
  \let\onelinecomment=\empty
  \let\commentbegin=\empty
  \let\commentend=\empty}

\visiblecomments

\newlength{\blanklineskip}
\newcommand{\hsindent}[1]{\quad}
\let\hspre\empty
\let\hspost\empty

\EndFmtInput
\makeatother
\ReadOnlyOnce{polycode.fmt}%
\makeatletter

\newcommand{\hsnewpar}[1]%
  {{\parskip=0pt\parindent=0pt\par\vskip #1\noindent}}

\newcommand{\hscodestyle}{}

\newcommand{\sethscode}[1]%
  {\expandafter\let\expandafter\hscode\csname #1\endcsname
   \expandafter\let\expandafter\endhscode\csname end#1\endcsname}

  {\par\noindent
   \advance\leftskip\mathindent
   \hscodestyle
   \let\\=\@normalcr
   \let\hspre\(\let\hspost\)%
   \pboxed}%
  {\endpboxed\)%
   \par\noindent
   \ignorespacesafterend}

  {\hsnewpar\abovedisplayskip
   \advance\leftskip\mathindent
   \hscodestyle
   \let\hspre\(\let\hspost\)%
   \pboxed}%
  {\endpboxed%
   \hsnewpar\belowdisplayskip
   \ignorespacesafterend}

  {\hsnewpar\abovedisplayskip
   \advance\leftskip\mathindent
   \hscodestyle
   \let\\=\@normalcr
   \(\pboxed}%
  {\endpboxed\)%
   \hsnewpar\belowdisplayskip
   \ignorespacesafterend}

\newcommand{\plainhs}{\sethscode{plainhscode}}

\plainhs

  {\hsnewpar\abovedisplayskip
   \advance\leftskip\mathindent
   \hscodestyle
   \let\\=\@normalcr
   \(\parray}%
  {\endparray\)%
   \hsnewpar\belowdisplayskip
   \ignorespacesafterend}

  {\parray}{\endparray}

  {\(\parray}{\endparray\)}

\def\codeframewidth{\arrayrulewidth}
\RequirePackage{calc}

  {\parskip=\abovedisplayskip\par\noindent
   \hscodestyle
   \arrayrulewidth=\codeframewidth
   \tabular{@{}|p{\linewidth-2\arraycolsep-2\arrayrulewidth-2pt}|@{}}%
   \hline\framedhslinecorrect\\{-1.5ex}%
   \let\endoflinesave=\\
   \let\\=\@normalcr
   \(\pboxed}%
  {\endpboxed\)%
   \framedhslinecorrect\endoflinesave{.5ex}\hline
   \endtabular
   \parskip=\belowdisplayskip\par\noindent
   \ignorespacesafterend}

\newcommand{\framedhslinecorrect}[2]%
  {#1[#2]}

  {\(\def\column##1##2{}%
   \let\>\undefined\let\<\undefined\let\\\undefined
   \newcommand\>[1][]{}\newcommand\<[1][]{}\newcommand\\[1][]{}%
   \def\fromto##1##2##3{##3}%
   }{\) }%

  {\let\orighscode=\hscode
   \let\origendhscode=\endhscode
   \def\endhscode{\def\hscode{\endgroup\def\@currenvir{hscode}\\}\begingroup}
   \orighscode\def\hscode{\endgroup\def\@currenvir{hscode}}}%
  {\origendhscode
   \global\let\hscode=\orighscode
   \global\let\endhscode=\origendhscode}%

\makeatother
\EndFmtInput
\def\choice#1{\mathbin{_{#1}\diamond}}

\def\wider#1{~ #1 ~}
\def\meither#1#2{\left[#1|#2\right]}        
\def\msplit#1#2{\left[\frac{#1}{#2}\right]} 

\def\rcb#1#2#3#4{\def\nothing{}\def\range{#3}\mathopen{\langle}#1 \ #2 \ \ifx\range\nothing::\else: \ #3 :\fi \ #4\mathclose{\rangle}}
\def\comp{\mathbin{\cdot}}
\def\conj#1#2{\mathopen{\langle} #1, #2 \mathclose{\rangle}}
\def\conv#1{#1^\circ}

\def\just#1#2{\\ &#1& \rule{2em}{0pt} \{ \mbox{\rule[-.7em]{0pt}{1.8em} \small #2 \/} \} \nonumber\\ && }
\def\longjust#1#2{\\ &#1& \rule{2em}{0pt}
\left\{\begin{tabular}{l}#2\end{tabular} \right\} \nonumber\\ && }
\def\implied{\mathbin\Leftarrow}

\begin{document}

\begin{abstract}
In the trend towards tolerating hardware unreliability, \emph{accuracy} is
exchanged for \emph{cost savings}. Running on less reliable machines, ``functionally
correct'' code becomes risky and one needs to know how risk propagates
so as to mitigate it.

Risk estimation, however,  seems to live outside
the average programmer's technical competence and core practice.

In this paper we propose that risk be constructively handled in
\emph{functional programming} by (a) writing programs which may choose between
\emph{expected} and \emph{faulty} behaviour,
and by (b) reasoning about them 
in a linear algebra extension to standard, {\em\`a la} Bird-Moor algebra of programming.

In particular, the
propagation of faults across standard program transformation techniques known as \emph{tupling} and
\emph{fusion} is calculated, enabling the \emph{fault of the whole} to be expressed
in terms of the \emph{faults of its parts}.
\end{abstract}

\maketitle

\section{Introduction}

With software so invasive in every-day's life as it is today, you don't need to
be staff of a space agency to place the question:
\emph{what {risks} do we run day-to-day by relying on so much software?}
\cite{Ja09} writes:
\begin{quote}\em
(...) a {dependable system} is one (..) in which you can place your reliance
or trust. A rational person or organization only does this with {evidence}
that the system's {benefits} outweigh its {risks}.
\end{quote}
Over the years,
NASA has defined a \emph{probabilistic risk assessment} (PRA) methodology
to enhance the safety decision process.
Quoting 
\citep{SD11}:
\begin{quote}\em
PRA characterizes risk in terms of three basic questions:
(1) What can \emph{go wrong}?
(2) How \emph{likely} is it? and
(3) What are the \emph{consequences}?
The PRA process
answers these questions by systematically
(...) identifying, modeling, and \emph{quantifying} scenarios that can
lead to undesired consequences.
\end{quote}
This may leave one with the feeling that PRA takes place
\emph{a posteriori}, that is, once the system is built. Even if
a wrong understanding of 
PRA, limitations of current programming
practice are apparent concerning timely assessment of the risks involved
in the future use of computer programs. 
\emph{Things that can go wrong} can be guessed; but, how is the \emph{likelihood} of such
bad behaviour expressed? and how does one quantify its \emph{consequences}
(fault propagation)?

This paper addresses these questions and issues in the context of \emph{functional programming}
(FP) over \emph{unreliable} hardware.
Note that such unreliability can be intentional, as is the case in \emph{inexact
circuit design} \citep{LEPP13}, where accuracy of the circuit is exchanged for cost
savings (eg.\ energy, delay, silicon). 

We will show that FP is well prepared for smoothly incorporating risk analysis
in the design of programs. This
is because the standard \emph{qualitative} semantics of FPs can evolve towards
a \emph{quantitative} one simply by upgrading its underlying \emph{relational} algebra 
of programs ``\`a la Bird-Moor'' (\citeyear{BM97}) into a \emph{linear} algebra of programming \citep{Ol12a}.

The basic idea is simple: suppose one writes function \ensuremath{\Varid{good}} for the intended
behaviour of a program and there is evidence that, with probability \ensuremath{\Varid{p}},
such behaviour can turn into a \ensuremath{\Varid{bad}} function.
Using the \emph{probabilistic choice} combinator \ensuremath{({\cdot }\choice{\cdot }{\cdot })} of \cite{MM05,Ol12a},
one may write term
\begin{quote}
\ensuremath{{\Varid{bad}}\choice{\Varid{p}}{\Varid{good}}}
\end{quote}
to express the complete (ie.\ with risk incorporated) behaviour of what one is programming.

What is needed, then, is a method for evaluating the
propagation of risk, for instance across recursion schemes.
This is what the \emph{linear} algebra of programming (LAoP) 
is intended for.
This paper investigates, in particular, the quantitative extension
of the so-called \emph{mutual recursion} and \emph{banana-split} laws \citep{BM97}
which underpin the refinement of primitive recursive
functions into linear implementations and checks
under what conditions are such implementations as good as their original
definitions with respect to fault propagation.

The approach will be illustrated in two ways: either
by running programs as probabilistic (monadic) functions written in Haskell
using the PFP library of \cite{EK06},
or by running finite approximations of them directly as
matrices in \matlab~\footnote{\matlab\ \texttrademark\ is a trademark of The MathWorks \textregistered.}.

\paragraph{Contribution}
In the trend towards tolerating hardware unreliability, \emph{accuracy} is exchang-ed for \emph{cost savings}. Running on less reliable machines, functionally ``correct'' code becomes risky and one needs to know how risk propagates so as to mitigate it.
In this context, this paper presents the following contributions:
\begin{itemize}
\item	
It shows how the standard \emph{algebra
of functional programs} dear to the so-called \emph{program transformation} school of software design 
extends and incorporates risk
simply by switching from \emph{``sharp''} functions to \emph{probabilistic} functions
handled as matrices in {linear algebra}.\footnote{This extends
to deterministic imperative programs via probabilistic functional semantics denotation.}
\item
The laws of such a \emph{linear} algebra of programming are shown to capture the notion of
probabilistic indistinguishability, essential to decide whether \emph{program transformation} rules
can be safely applied or not.
\item
The approach is shown to be readily applicable to \emph{recursive
programs} which handle possibly interfering threads of computation.
\item
In particular, mutually recursive computations are addressed showing under what conditions
mutual recursion slicing holds in the probabilistic setting.
\item
Finally, the paper shows that a well-known \emph{tupling} technique known as the \emph{``banana-split''}
functional program transformation is still valid in presence of faults.
\end{itemize}

\paragraph{Paper outline}
The following section presents two motivating programs which
will be subject to fault-injection as an illustration of risk simulation
and calculation. Section \ref{sec:130320a} addresses the derivation of such
programs via mutual-recursion transformation, an exercise which is extended in section
\ref{sec:130324b} to the probabilistic setting. 
A basis for this is given in section \ref{sec:130326a},
where the LAoP 
is put in context, leading to the
approach to probabilistic mutual recursion given in section \ref{sec:130326b}.
This in turn leads to an
asymmetry (section \ref{sec:130326c}) which explains the different fault
propagation patterns found in the two motivating examples (section \ref{sec:130326d}).
The topic of fault propagation in functional programming is further delved
in section \ref{sec:130326e} by moving to more elaborate data types and
showing how the \emph{risk of the whole} can be calculated combining the
\emph{risk of the parts}. The two last sections conclude, review related work and give prospects for
future work.  Proofs of auxiliary results are deferred to appendix \ref{sec:130324a}.

\section{Motivation}
\label{sec:130320b}
Let us start from two programs written in C, one
which supposedly computes the square of a non-negative integer \ensuremath{\Varid{n}},
\begin{quote}\small
\begin{tabbing}\tt
~int~sq\char40{}int~n\char41{}~\char123{}\\
\tt ~~~int~s\char61{}0\char59{}~int~o\char61{}1\char59{}\\
\tt ~~~int~i\char59{}\\
\tt ~~~for~\char40{}i\char61{}1\char59{}i\char60{}n\char43{}1\char59{}i\char43{}\char43{}\char41{}~\char123{}s\char43{}\char61{}o\char59{}~o\char43{}\char61{}2\char59{}\char125{}\\
\tt ~~~return~s\char59{}\\
\tt ~\char125{}\char59{}
\end{tabbing}
\end{quote}
and the other
\begin{quote}\small
\begin{tabbing}\tt
~int~fib\char40{}int~n\char41{}~\char123{}\\
\tt ~~~int~x\char61{}0\char59{}~int~y\char61{}1\char59{}~int~i\char59{}\\
\tt ~~~for~\char40{}i\char61{}1\char59{}i\char60{}\char61{}n\char59{}i\char43{}\char43{}\char41{}~\char123{}int~a\char61{}y\char59{}~y\char61{}y\char43{}x\char59{}~x\char61{}a\char59{}\char125{}\\
\tt ~~~return~x\char59{}\\
\tt ~\char125{}\char59{}
\end{tabbing}
\end{quote}
which supposedly computes the \ensuremath{\Varid{n}}-th entry in the Fibonacci series,
for \ensuremath{\Varid{n}} positive.

Both programs are \ensuremath{\mathsf{for}}-loops whose bodies rely on the same operation: addition of
natural numbers. Suppose one knows that,
in the machine where such programs will run,
there is the risk of addition misbehaving in some known way: with probability \ensuremath{\Varid{p}},
\ensuremath{\Varid{x}\mathbin{+}\Varid{y}} may evaluate to \ensuremath{\Varid{y}}, in which case \ensuremath{(\Varid{x}\mathbin{+})\mathrel{=}\Varid{id}}, the identity function.
Or one might know that, in some unfriendly environment, the processor's arithmetic-logic
unit may reset addition output to \ensuremath{\mathrm{0}}, with probability \ensuremath{\Varid{q}}.

The question is: what is the impact of such faults in the overall behaviour
of each \ensuremath{\mathsf{for}}-loop? Can we \emph{measure} such an impact? Can we \emph{predict} it?
Are there versions of the same programs which mitigate such faults better
than the ones given?

The standard approach to these questions relies on simulation:
one performs a large number of experiments in which the programs
run with the given \emph{faults injected} according to the given probabilities
and then performs statistic analysis of the outcome of such simulations.
Software \emph{fault injection} \citep{VM97} is a
more and more widespread technique for quality assurance which measures
the propagation of faults through paths that might otherwise rarely be followed
in testing. The G-SWFIT technique, for instance, emulates the software
fault classes most frequently observed in the field through a library of fault emulation operators,
and injects such faults directly in the target executable code \citep{DM06}.

In this paper we adopt a different strategy: instead of simulating risky behaviour
\emph{a posteriori},
this is taken into account \emph{a priori} by moving from imperative to functional code
whereby faulty behaviour is encoded in terms of probabilistic functions \citep{EK06}.
Take the two versions of faulty addition given above as examples:
the first can be expressed by turning \ensuremath{(\mathbin{+})} into the probabilistic function
\begin{quote}
\ensuremath{\mathit{fadd}_{\Varid{x}}\mathrel{=}{\Varid{id}}\choice{\Varid{p}}{(\Varid{x}\mathbin{+})}}
\end{quote}
(\ensuremath{\mathit{fadd}_{\cdot }} for ``faulty addition'') which misbehaves as the identity function \ensuremath{\Varid{id}} with probability \ensuremath{\Varid{p}} and exhibits
the correct behaviour with probability \ensuremath{\mathrm{1}\mathbin{-}\Varid{p}}; similarly,
the second version is expressed by probabilistic choice
\begin{quote}
\ensuremath{\mathit{fadd}_{\Varid{x}}\mathrel{=}{\underline{\mathrm{0}}}\choice{\Varid{q}}{(\Varid{x}\mathbin{+})}}
\end{quote}
where \ensuremath{\underline{\mathrm{0}}\;\anonymous \mathrel{=}\mathrm{0}} is the everywhere-\ensuremath{\mathrm{0}} constant function.
Of course, we might think of more elaborate fault patterns, for instance
\begin{quote}
\ensuremath{\mathit{fadd}_{\Varid{x}}\mathrel{=}{({\underline{\mathrm{0}}}\choice{\Varid{q}}{\Varid{id}})}\choice{\Varid{p}}{(\Varid{x}\mathbin{+})}}
\end{quote}
in which the probability of \ensuremath{\mathit{fadd}_{\cdot }} resetting to \ensuremath{\mathrm{0}} is \ensuremath{\Varid{qp}} and \ensuremath{(\mathrm{1}\mathbin{-}\Varid{q})\;\Varid{p}} is
that of degenerating into the identity; or even thinking of normal
distributions centered upon the expected output \ensuremath{\Varid{x}\mathbin{+}\Varid{y}}, and so on.

Probabilistic functions are distribution-valued functions which can be written
in the monadic style over the \emph{distribution monad}. This is termed
\ensuremath{\Conid{Dist}} in the PFP library of \cite{EK06}, which we shall be using in the sequel.%
\footnote{All distributions in our approach are generated by finite
application of the \ensuremath{\Varid{choice}} operator and therefore have finite support.}
Moreover, probabilistic functions can be reasoned about using the laws of monads,
explicitly as advocated by \cite{GH11} or implicitly as in the probabilistic notation
proposed by \cite{Mor12} as extension to the standard 
Eindhoven quantifier calculus \citep{BM06}.
 
There is yet another alternative: every probabilistic function \ensuremath{\Varid{f}\mathbin{:}\Varid{a}\to \Conid{Dist}\;\Varid{b}} is in one-to-one correspondence with a matrix whose columns are indexed
by \ensuremath{\Varid{a}}, whose rows are indexed by \ensuremath{\Varid{b}} and whose multiplication corresponds
to composition in the Kleisli category induced by \ensuremath{\Conid{Dist}}
\citep{Ol12a,Ol12b}. This offers the possibility of using the rich field of
\emph{linear algebra} to calculate with probabilistic functions,
in the same way relation algebra is advocated by \cite{BM97}  for
reasoning about standard (sharp) functions.

One of the advantages of such a \emph{linear} algebra of programming (LAoP)
is the way recursive probabilistic functions are handled --- simply by
using the same combinators (eg.\ maps, folds) --- of the standard
algebra of programming \citep{BM97}. The shift from a qualitative to a quantitative
semantics is therefore rather smooth --- the game is the same, the move
ensured just by change of underlying category.
Following this approach, \cite{Ol12a} already gives an example of what might
be referred to as \emph{fault-fusion}: the risk of the whole 
misbehaving can be expressed in terms of the risk of the parts misbehaving
wherever a particular fusion law is applicable.

Note, however, that not every law of the algebra of programming extends quantitatively.
 In this paper we address the linear algebra extension of one such law which
is particularly relevant to program calculation: the \emph{mutual
recursion} 
law enabling systems of mutually recursive functions
to be merged into a single, more efficient function \citep{BM97}. Both C programs given
above can be derived from their specifications using such a law. Below
we show how they can be turned into probabilistic functions expressing
safe and risky behaviour in a natural and calculational way.

\section{Mutual recursion}
\label{sec:130320a}
Let us take the standard definition of the Fibonacci function, written in Haskell syntax:
\begin{hscode}\SaveRestoreHook
\column{B}{@{}>{\hspre}l<{\hspost}@{}}%
\column{E}{@{}>{\hspre}l<{\hspost}@{}}%
\>[B]{}\Varid{fib}\;\mathrm{0}\mathrel{=}\mathrm{0}{}\<[E]%
\\
\>[B]{}\Varid{fib}\;\mathrm{1}\mathrel{=}\mathrm{1}{}\<[E]%
\\
\>[B]{}\Varid{fib}\;(\Varid{n}\mathbin{+}\mathrm{2})\mathrel{=}\Varid{fib}\;\Varid{n}\mathbin{+}\Varid{fib}\;(\Varid{n}\mathbin{+}\mathrm{1}){}\<[E]%
\ColumnHook
\end{hscode}\resethooks
The linear version encoded in the C program given above is obtained by
pairing \ensuremath{\Varid{fib}} with its \emph{derivative}, \ensuremath{\Varid{f}\;\Varid{n}\mathrel{=}\Varid{fib}\;(\Varid{n}\mathbin{+}\mathrm{1})}:
\footnote{Since \ensuremath{\Varid{f}\;\mathrm{0}\mathrel{=}\Varid{fib}\;\mathrm{1}\mathrel{=}\mathrm{1}} and \ensuremath{\Varid{f}\;(\Varid{n}\mathbin{+}\mathrm{1})\mathrel{=}\Varid{fib}\;(\Varid{n}\mathbin{+}\mathrm{2})\mathrel{=}\Varid{fib}\;\Varid{n}\mathbin{+}\Varid{fib}\;(\Varid{n}\mathbin{+}\mathrm{1})\mathrel{=}\Varid{fib}\;\Varid{n}\mathbin{+}\Varid{f}\;\Varid{n}}.}
\begin{hscode}\SaveRestoreHook
\column{B}{@{}>{\hspre}l<{\hspost}@{}}%
\column{E}{@{}>{\hspre}l<{\hspost}@{}}%
\>[B]{}\Varid{fib}\;\mathrm{0}\mathrel{=}\mathrm{0}{}\<[E]%
\\
\>[B]{}\Varid{fib}\;(\Varid{n}\mathbin{+}\mathrm{1})\mathrel{=}\Varid{f}\;\Varid{n}{}\<[E]%
\\[\blanklineskip]%
\>[B]{}\Varid{f}\;\mathrm{0}\mathrel{=}\mathrm{1}{}\<[E]%
\\
\>[B]{}\Varid{f}\;(\Varid{n}\mathbin{+}\mathrm{1})\mathrel{=}\Varid{fib}\;\Varid{n}\mathbin{+}\Varid{f}\;\Varid{n}{}\<[E]%
\ColumnHook
\end{hscode}\resethooks
The pairing of the two functions,
\begin{hscode}\SaveRestoreHook
\column{B}{@{}>{\hspre}l<{\hspost}@{}}%
\column{E}{@{}>{\hspre}l<{\hspost}@{}}%
\>[B]{}(\Varid{fib},\Varid{f})\;\Varid{n}\mathrel{=}(\Varid{fib}\;\Varid{n},\Varid{f}\;\Varid{n}){}\<[E]%
\ColumnHook
\end{hscode}\resethooks
can be expressed primitive-recursively by
\begin{hscode}\SaveRestoreHook
\column{B}{@{}>{\hspre}l<{\hspost}@{}}%
\column{E}{@{}>{\hspre}l<{\hspost}@{}}%
\>[B]{}(\Varid{fib},\Varid{f})\;\mathrm{0}\mathrel{=}(\Varid{fib}\;\mathrm{0},\Varid{f}\;\mathrm{0})\mathrel{=}(\mathrm{0},\mathrm{1}){}\<[E]%
\\
\>[B]{}(\Varid{fib},\Varid{f})\;(\Varid{n}\mathbin{+}\mathrm{1})\mathrel{=}(\Varid{f}\;\Varid{n},\Varid{fib}\;\Varid{n}\mathbin{+}\Varid{f}\;\Varid{n}){}\<[E]%
\ColumnHook
\end{hscode}\resethooks
or by the equivalent
\begin{hscode}\SaveRestoreHook
\column{B}{@{}>{\hspre}l<{\hspost}@{}}%
\column{4}{@{}>{\hspre}l<{\hspost}@{}}%
\column{31}{@{}>{\hspre}l<{\hspost}@{}}%
\column{E}{@{}>{\hspre}l<{\hspost}@{}}%
\>[4]{}(\Varid{fib},\Varid{f})\;\mathrm{0}\mathrel{=}(\mathrm{0},\mathrm{1}){}\<[E]%
\\
\>[4]{}(\Varid{fib},\Varid{f})\;(\Varid{n}\mathbin{+}\mathrm{1})\mathrel{=}(\Varid{y},\Varid{x}\mathbin{+}\Varid{y})\;{}\<[31]%
\>[31]{}\mathbf{where}\;(\Varid{x},\Varid{y})\mathrel{=}(\Varid{fib},\Varid{f})\;\Varid{n}{}\<[E]%
\ColumnHook
\end{hscode}\resethooks
itself the same as
\begin{hscode}\SaveRestoreHook
\column{B}{@{}>{\hspre}l<{\hspost}@{}}%
\column{4}{@{}>{\hspre}l<{\hspost}@{}}%
\column{10}{@{}>{\hspre}l<{\hspost}@{}}%
\column{E}{@{}>{\hspre}l<{\hspost}@{}}%
\>[4]{}(\Varid{fib},\Varid{f})\mathrel{=}\mathsf{for}\;\Varid{loop}\;(\mathrm{0},\mathrm{1}){}\<[E]%
\\
\>[4]{}\hsindent{6}{}\<[10]%
\>[10]{}\mathbf{where}\;\Varid{loop}\;(\Varid{x},\Varid{y})\mathrel{=}(\Varid{y},\Varid{x}\mathbin{+}\Varid{y}){}\<[E]%
\ColumnHook
\end{hscode}\resethooks
by introduction of the \ensuremath{\mathsf{for}\;\Varid{loop}} combinator,
\begin{hscode}\SaveRestoreHook
\column{B}{@{}>{\hspre}l<{\hspost}@{}}%
\column{E}{@{}>{\hspre}l<{\hspost}@{}}%
\>[B]{}\mathsf{for}\;\Varid{b}\;\Varid{i}\;\mathrm{0}\mathrel{=}\Varid{i}{}\<[E]%
\\
\>[B]{}\mathsf{for}\;\Varid{b}\;\Varid{i}\;(\Varid{n}\mathbin{+}\mathrm{1})\mathrel{=}\Varid{b}\;(\mathsf{for}\;\Varid{b}\;\Varid{i}\;\Varid{n}){}\<[E]%
\ColumnHook
\end{hscode}\resethooks
where \ensuremath{\Varid{b}} is the loop body and \ensuremath{\Varid{i}} provides for initialization.
This is the natural-number equivalent to combinator \ensuremath{\mathsf{foldr}} over finite lists
in Haskell, ie.\ the \emph{catamorphism} \citep{BM97} \emph{of the natural numbers}.
Therefore, we can define
\begin{hscode}\SaveRestoreHook
\column{B}{@{}>{\hspre}l<{\hspost}@{}}%
\column{5}{@{}>{\hspre}l<{\hspost}@{}}%
\column{9}{@{}>{\hspre}l<{\hspost}@{}}%
\column{E}{@{}>{\hspre}l<{\hspost}@{}}%
\>[B]{}\Varid{fibl}\;\Varid{n}\mathrel{=}{}\<[E]%
\\
\>[B]{}\hsindent{5}{}\<[5]%
\>[5]{}\mathbf{let}\;(\Varid{x},\Varid{y})\mathrel{=}\mathsf{for}\;\Varid{loop}\;(\mathrm{0},\mathrm{1})\;\Varid{n}{}\<[E]%
\\
\>[5]{}\hsindent{4}{}\<[9]%
\>[9]{}\Varid{loop}\;(\Varid{x},\Varid{y})\mathrel{=}(\Varid{y},\Varid{x}\mathbin{+}\Varid{y}){}\<[E]%
\\
\>[B]{}\hsindent{5}{}\<[5]%
\>[5]{}\mathbf{in}\;\Varid{x}{}\<[E]%
\ColumnHook
\end{hscode}\resethooks
as the linear version of \ensuremath{\Varid{fib}} obtained by pairing \ensuremath{\Varid{fib}} with its derivative
--- compare with the C program given above.

The other program computing squares
can be derived in the same way from the specification \ensuremath{\Varid{sq}\;\Varid{n}\mathrel{=}{\Varid{n}}^{\mathrm{2}}}:
the two mutually recursive functions
\begin{hscode}\SaveRestoreHook
\column{B}{@{}>{\hspre}l<{\hspost}@{}}%
\column{E}{@{}>{\hspre}l<{\hspost}@{}}%
\>[B]{}\Varid{sq}\;\mathrm{0}\mathrel{=}\mathrm{0}{}\<[E]%
\\
\>[B]{}\Varid{sq}\;(\Varid{n}\mathbin{+}\mathrm{1})\mathrel{=}\Varid{sq}\;\Varid{n}\mathbin{+}\Varid{odd}\;\Varid{n}{}\<[E]%
\\[\blanklineskip]%
\>[B]{}\Varid{odd}\;\mathrm{0}\mathrel{=}\mathrm{1}{}\<[E]%
\\
\>[B]{}\Varid{odd}\;(\Varid{n}\mathbin{+}\mathrm{1})\mathrel{=}\mathrm{2}\mathbin{+}\Varid{odd}\;\Varid{n}{}\<[E]%
\ColumnHook
\end{hscode}\resethooks
arise from the binomial $(n+1)^2 = n^2 + 2n + 1$ and introduction of function \ensuremath{\Varid{odd}\;\Varid{n}\mathrel{=}\mathrm{2}\;\Varid{n}\mathbin{+}\mathrm{1}}, thus named because \ensuremath{\mathrm{2}\;\Varid{n}\mathbin{+}\mathrm{1}} is the $n$-th odd number.
(That is, the square of a natural number always is a sum of odd numbers.)
Pairing them up into \ensuremath{(\Varid{sq},\Varid{odd})\;\Varid{x}\mathrel{=}(\Varid{sq}\;\Varid{x},\Varid{odd}\;\Varid{x})} and proceeding in the same way
as above we obtain \ensuremath{(\Varid{sq},\Varid{odd})\mathrel{=}\mathsf{for}\;\Varid{loop}\;(\mathrm{0},\mathrm{1})} where \ensuremath{\Varid{loop}\;(\Varid{s},\Varid{o})\mathrel{=}(\Varid{s}\mathbin{+}\Varid{o},\Varid{o}\mathbin{+}\mathrm{2})}
and thereupon the following functional version of the given C program:
\footnote{Notice how the syntax \texttt{s+=o; o+=2;} in C nicely tallies with
\ensuremath{(\Varid{s}\mathbin{+}\Varid{o},\Varid{o}\mathbin{+}\mathrm{2})} in Haskell.}
\begin{hscode}\SaveRestoreHook
\column{B}{@{}>{\hspre}l<{\hspost}@{}}%
\column{4}{@{}>{\hspre}l<{\hspost}@{}}%
\column{5}{@{}>{\hspre}l<{\hspost}@{}}%
\column{8}{@{}>{\hspre}l<{\hspost}@{}}%
\column{E}{@{}>{\hspre}l<{\hspost}@{}}%
\>[B]{}\Varid{sql}\;\Varid{n}\mathrel{=}{}\<[E]%
\\
\>[B]{}\hsindent{4}{}\<[4]%
\>[4]{}\mathbf{let}\;(\Varid{s},\Varid{o})\mathrel{=}\mathsf{for}\;\Varid{loop}\;(\mathrm{0},\mathrm{1})\;\Varid{n}{}\<[E]%
\\
\>[4]{}\hsindent{4}{}\<[8]%
\>[8]{}\Varid{loop}\;(\Varid{s},\Varid{o})\mathrel{=}(\Varid{s}\mathbin{+}\Varid{o},\Varid{o}\mathbin{+}\mathrm{2}){}\<[E]%
\\
\>[4]{}\hsindent{1}{}\<[5]%
\>[5]{}\mathbf{in}\;\Varid{s}{}\<[E]%
\ColumnHook
\end{hscode}\resethooks
Clearly, each recursive function above and its linear version are,
extensionally, the same function. Let us now see what happens
once we start injecting risky (faulty) behaviour in each of them.

\section{Going probabilistic}
\label{sec:130324b}
Probabilistic extensions of any of the functions above can be obtained
by writing them monadically and then instantiating them
with the distribution monad \citep{EK06}.
Take the recursive version of \ensuremath{\Varid{fib}} given in the beginning of section \ref{sec:130320a}
and ``monadify it'' into:
\begin{hscode}\SaveRestoreHook
\column{B}{@{}>{\hspre}l<{\hspost}@{}}%
\column{3}{@{}>{\hspre}l<{\hspost}@{}}%
\column{14}{@{}>{\hspre}l<{\hspost}@{}}%
\column{E}{@{}>{\hspre}l<{\hspost}@{}}%
\>[B]{}\Varid{mfib}\;\mathrm{0}\mathrel{=}\Varid{return}\;\mathrm{0}{}\<[E]%
\\
\>[B]{}\Varid{mfib}\;\mathrm{1}\mathrel{=}\Varid{return}\;\mathrm{1}{}\<[E]%
\\
\>[B]{}\Varid{mfib}\;(\Varid{n}\mathbin{+}\mathrm{2})\mathrel{=}{}\<[E]%
\\
\>[B]{}\hsindent{3}{}\<[3]%
\>[3]{}\mathbf{do}\;\{\mskip1.5mu \Varid{x}\leftarrow {}\<[14]%
\>[14]{}\Varid{mfib}\;\Varid{n};\Varid{y}\leftarrow \Varid{mfib}\;(\Varid{n}\mathbin{+}\mathrm{1});\Varid{return}\;(\Varid{x}\mathbin{+}\Varid{y})\mskip1.5mu\}{}\<[E]%
\ColumnHook
\end{hscode}\resethooks
Running \ensuremath{\Varid{mfib}\;\Varid{n}} inside the \ensuremath{\Conid{Dist}} monad one gets \ensuremath{\Varid{fib}\;\Varid{n}} with \ensuremath{\mathrm{100}\mathbin{\%}} probability,
since \ensuremath{\Varid{return}} yields the \emph{one-point}, Dirac distribution of its argument.

Now let us inject one of the faults mentioned in section \ref{sec:130320b},
say \ensuremath{\mathit{fadd}_{\Varid{p}}\;\Varid{x}\mathrel{=}{\Varid{id}}\choice{\Varid{p}}{(\Varid{x}\mathbin{+})}} with \ensuremath{\Varid{p}\mathrel{=}\mathrm{0.1}}, for instance. For this we just replace
\ensuremath{\Varid{return}\;(\Varid{x}\mathbin{+}\Varid{y})} (perfect addition) by \ensuremath{\mathit{fadd}_{\mathrm{0.1}}\;\Varid{x}\;\Varid{y}} and run test cases, eg.
\footnote{The probabilities in this example and others to follow
are chosen with no criterion at all apart from leading to 
distributions visible to the naked eye. By all means, \ensuremath{\mathrm{0.1}} would be extremely
high risk in realistic PRA \citep{SD11}, where only figures as small as \ensuremath{\mathrm{1.0}}E-\ensuremath{\mathrm{7}} are
``acceptable'' risks.}
\begin{quote}\small
\begin{tabbing}\tt
~Main\char62{}~mfib~4\\
\tt ~3~~81\char46{}0\char37{}\\
\tt ~2~~18\char46{}0\char37{}\\
\tt ~1~~~1\char46{}0\char37{}
\end{tabbing}
\end{quote}
We see that the correct behaviour (\ensuremath{\mathrm{100}\mathbin{\%}} chance of getting \ensuremath{\Varid{fib}\;\mathrm{4}\mathrel{=}\mathrm{3}})
is no longer ensured --- with chance \ensuremath{\mathrm{18}\mathbin{\%}} one may get \ensuremath{\mathrm{2}} as result and even \ensuremath{\mathrm{1}} is a
possible output, with probability \ensuremath{\mathrm{1}\mathbin{\%}}.

Similar experiments can be carried out with the linear version by defining its monadic evolution
\begin{hscode}\SaveRestoreHook
\column{B}{@{}>{\hspre}l<{\hspost}@{}}%
\column{5}{@{}>{\hspre}l<{\hspost}@{}}%
\column{20}{@{}>{\hspre}l<{\hspost}@{}}%
\column{E}{@{}>{\hspre}l<{\hspost}@{}}%
\>[B]{}\Varid{mfibl}\;\Varid{n}\mathrel{=}{}\<[E]%
\\
\>[B]{}\hsindent{5}{}\<[5]%
\>[5]{}\mathbf{do}\;\{\mskip1.5mu (\Varid{x},\Varid{y})\leftarrow {}\<[20]%
\>[20]{}\Varid{mfor}\;\Varid{loop}\;(\mathrm{0},\mathrm{1})\;\Varid{n};\Varid{return}\;\Varid{x}\mskip1.5mu\}{}\<[E]%
\\
\>[B]{}\hsindent{5}{}\<[5]%
\>[5]{}\mathbf{where}\;\Varid{loop}\;(\Varid{x},\Varid{y})\mathrel{=}\Varid{return}\;(\Varid{y},\Varid{x}\mathbin{+}\Varid{y}){}\<[E]%
\ColumnHook
\end{hscode}\resethooks
relying on the monadic extension of the \ensuremath{\mathsf{for}} combinator:
\begin{hscode}\SaveRestoreHook
\column{B}{@{}>{\hspre}l<{\hspost}@{}}%
\column{E}{@{}>{\hspre}l<{\hspost}@{}}%
\>[B]{}\Varid{mfor}\;\Varid{b}\;\Varid{i}\;\mathrm{0}\mathrel{=}\Varid{return}\;\Varid{i}{}\<[E]%
\\
\>[B]{}\Varid{mfor}\;\Varid{b}\;\Varid{i}\;(\Varid{n}\mathbin{+}\mathrm{1})\mathrel{=}\mathbf{do}\;\{\mskip1.5mu \Varid{x}\leftarrow \Varid{mfor}\;\Varid{b}\;\Varid{i}\;\Varid{n};\Varid{b}\;\Varid{x}\mskip1.5mu\}{}\<[E]%
\ColumnHook
\end{hscode}\resethooks
To inject into \ensuremath{\Varid{mfibl}} the same fault injected before into \ensuremath{\Varid{mfib}} amounts to
replacing, in the loop body, \emph{good} addition by the \emph{bad} one:
\begin{hscode}\SaveRestoreHook
\column{B}{@{}>{\hspre}l<{\hspost}@{}}%
\column{E}{@{}>{\hspre}l<{\hspost}@{}}%
\>[B]{}\Varid{loop}\;(\Varid{x},\Varid{y})\mathrel{=}\mathbf{do}\;\{\mskip1.5mu \Varid{z}\leftarrow \mathit{fadd}_{\mathrm{0.1}}\;\Varid{x}\;\Varid{y};\Varid{return}\;(\Varid{y},\Varid{z})\mskip1.5mu\}{}\<[E]%
\ColumnHook
\end{hscode}\resethooks
Running the same experiment as above we still get \ensuremath{\Varid{mfibl}\;\mathrm{4}\mathrel{=}\Varid{mfib}\;\mathrm{4}}.
However, behavioural equality between the two
(one recursive, the other linear) fault-injected
versions of \ensuremath{\Varid{fib}} is no longer true for arguments \ensuremath{\Varid{n}\mathbin{>}\mathrm{4}}, see for
instance
\begin{quote}\small
\begin{tabular}{c|c|c}
\ensuremath{\Varid{n}} & \ensuremath{\Varid{mfib}\;\Varid{n}} & \ensuremath{\Varid{mfibl}\;\Varid{n}}
\\\hline
	5
&
	\begin{minipage}{30em}
\begin{tabbing}\tt
~5~~65\char46{}6\char37{}\\
\tt ~4~~21\char46{}9\char37{}\\
\tt ~3~~10\char46{}5\char37{}\\
\tt ~2~~~1\char46{}9\char37{}\\
\tt ~1~~~0\char46{}1\char37{}
\end{tabbing}
	\end{minipage}
&
	\begin{minipage}{30em}
\begin{tabbing}\tt
~5~~72\char46{}9\char37{}\\
\tt ~3~~16\char46{}2\char37{}\\
\tt ~4~~~8\char46{}1\char37{}\\
\tt ~2~~~2\char46{}7\char37{}\\
\tt ~1~~~0\char46{}1\char37{}
\end{tabbing}
	\end{minipage}
\\\hline
	6
&
	\begin{minipage}{30em}
\begin{tabbing}\tt
~8~~47\char46{}8\char37{}\\
\tt ~7~~26\char46{}6\char37{}\\
\tt ~6~~11\char46{}8\char37{}\\
\tt ~5~~~9\char46{}8\char37{}\\
\tt ~4~~~2\char46{}7\char37{}\\
\tt ~3~~~1\char46{}1\char37{}\\
\tt ~2~~~0\char46{}2\char37{}\\
\tt ~1~~~0\char46{}0\char37{}
\end{tabbing}
	\end{minipage}
&
	\begin{minipage}{30em}
\begin{tabbing}\tt
~8~~65\char46{}6\char37{}\\
\tt ~6~~14\char46{}6\char37{}\\
\tt ~5~~14\char46{}6\char37{}\\
\tt ~3~~~2\char46{}4\char37{}\\
\tt ~4~~~2\char46{}4\char37{}\\
\tt ~2~~~0\char46{}4\char37{}\\
\tt ~1~~~0\char46{}0\char37{}
\end{tabbing}
	\end{minipage}
\end{tabular}
\end{quote}
the linear version performing better than the recursive one
in the sense of hitting the correct answer with higher 
probability.

\def\hiddenTodo{Important: to check this run \ensuremath{\Varid{mfib'}} \ensuremath{\Varid{mfibl'}} (where \ensuremath{\Varid{loop}} is free). Cf. Aux section}

Finally, let us now carry out similar experiments concerning the injection of the same fault
(in the addition function) in suitably extended (monadic) versions of
the square function, the recursive one
\begin{hscode}\SaveRestoreHook
\column{B}{@{}>{\hspre}l<{\hspost}@{}}%
\column{E}{@{}>{\hspre}l<{\hspost}@{}}%
\>[B]{}\Varid{msq}\;\mathrm{0}\mathrel{=}\Varid{return}\;\mathrm{0}{}\<[E]%
\\
\>[B]{}\Varid{msq}\;(\Varid{n}\mathbin{+}\mathrm{1})\mathrel{=}\mathbf{do}\;\{\mskip1.5mu \Varid{m}\leftarrow \Varid{msq}\;\Varid{n};\mathit{fadd}_{\mathrm{0.1}}\;\Varid{m}\;(\mathrm{2}\mathbin{*}\Varid{n}\mathbin{+}\mathrm{1})\mskip1.5mu\}{}\<[E]%
\ColumnHook
\end{hscode}\resethooks
and the linear one:
\begin{hscode}\SaveRestoreHook
\column{B}{@{}>{\hspre}l<{\hspost}@{}}%
\column{4}{@{}>{\hspre}l<{\hspost}@{}}%
\column{7}{@{}>{\hspre}l<{\hspost}@{}}%
\column{16}{@{}>{\hspre}l<{\hspost}@{}}%
\column{E}{@{}>{\hspre}l<{\hspost}@{}}%
\>[B]{}\Varid{msql}\;\Varid{n}\mathrel{=}{}\<[E]%
\\
\>[B]{}\hsindent{4}{}\<[4]%
\>[4]{}\mathbf{do}\;\{\mskip1.5mu (\Varid{s},\Varid{o})\leftarrow \Varid{mfor}\;\Varid{loop}\;(\mathrm{0},\mathrm{1})\;\Varid{n};\Varid{return}\;\Varid{s}\mskip1.5mu\}{}\<[E]%
\\
\>[4]{}\hsindent{3}{}\<[7]%
\>[7]{}\mathbf{where}\;\Varid{loop}\;(\Varid{s},\Varid{o})\mathrel{=}{}\<[E]%
\\
\>[7]{}\hsindent{9}{}\<[16]%
\>[16]{}\mathbf{do}\;\{\mskip1.5mu \Varid{z}\leftarrow \mathit{fadd}_{\mathrm{0.1}}\;\Varid{s}\;\Varid{o};\Varid{return}\;(\Varid{z},\Varid{o}\mathbin{+}\mathrm{2})\mskip1.5mu\}{}\<[E]%
\ColumnHook
\end{hscode}\resethooks
In this case --- as much as we can test --- both versions exhibit
the same behaviour, that is, they are probabilistically indistinguishable,
see for instance:
\begin{quote}\small
\begin{tabular}{c|c|c}
\ensuremath{\Varid{n}} & \ensuremath{\Varid{msq}\;\Varid{n}} & \ensuremath{\Varid{msql}\;\Varid{n}}
\\\hline
	0
&
	\begin{minipage}{30em}
\begin{tabbing}\tt
~0~100\char46{}0\char37{}
\end{tabbing}
	\end{minipage}
&
	\begin{minipage}{30em}
\begin{tabbing}\tt
~0~100\char46{}0\char37{}
\end{tabbing}
	\end{minipage}
\\\hline
	1
&
	\begin{minipage}{30em}
\begin{tabbing}\tt
~1~100\char46{}0\char37{}
\end{tabbing}
	\end{minipage}
&
	\begin{minipage}{30em}
\begin{tabbing}\tt
~1~100\char46{}0\char37{}
\end{tabbing}
	\end{minipage}
\\\hline
	2
&
	\begin{minipage}{30em}
\begin{tabbing}\tt
~4~~90\char46{}0\char37{}\\
\tt ~3~~10\char46{}0\char37{}
\end{tabbing}
	\end{minipage}
&
	\begin{minipage}{30em}
\begin{tabbing}\tt
~4~~90\char46{}0\char37{}\\
\tt ~3~~10\char46{}0\char37{}
\end{tabbing}
	\end{minipage}
\\\hline
	3
&
	\begin{minipage}{30em}
\begin{tabbing}\tt
~9~~81\char46{}0\char37{}\\
\tt ~5~~10\char46{}0\char37{}\\
\tt ~8~~~9\char46{}0\char37{}
\end{tabbing}
	\end{minipage}
&
	\begin{minipage}{30em}
\begin{tabbing}\tt
~9~~81\char46{}0\char37{}\\
\tt ~5~~10\char46{}0\char37{}\\
\tt ~8~~~9\char46{}0\char37{}
\end{tabbing}
	\end{minipage}
\\\hline
	\vdots
&
	\vdots
&
	\vdots
\\\hline
	6
&
	\begin{minipage}{30em}
\begin{tabbing}\tt
~36~~59\char46{}0\char37{}\\
\tt ~11~~10\char46{}0\char37{}\\
\tt ~20~~~9\char46{}0\char37{}\\
\tt ~27~~~8\char46{}1\char37{}\\
\tt ~32~~~7\char46{}3\char37{}\\
\tt ~35~~~6\char46{}6\char37{}
\end{tabbing}
	\end{minipage}
&
	\begin{minipage}{30em}
\begin{tabbing}\tt
~36~~59\char46{}0\char37{}\\
\tt ~11~~10\char46{}0\char37{}\\
\tt ~20~~~9\char46{}0\char37{}\\
\tt ~27~~~8\char46{}1\char37{}\\
\tt ~32~~~7\char46{}3\char37{}\\
\tt ~35~~~6\char46{}6\char37{}
\end{tabbing}
	\end{minipage}
\\\hline
	\vdots
&
	\vdots
&
	\vdots
\end{tabular}
\end{quote}

Summing up, we are in presence of two examples in which the risk of bad behaviour
propagates differently across the mutual recursion functional program transformation.

In the remainder of this paper we will resort to linear algebra to explain this
discrepancy.
We will show that, even if the transformation does not hold in general for
probabilistic functions, there are side conditions sufficient for it to hold,
explaining the different behaviour witnessed in the examples above.

\section{Probabilistic \ensuremath{\mathsf{for}}-loops in the LAoP}
\label{sec:130326a}
Consider the probabilistic Boolean function \ensuremath{\Varid{f}\mathrel{=}{\underline{\Conid{False}}}\choice{\mathrm{0.05}}{(\neg )}}
which is such that
\ensuremath{\Varid{f}\;\Conid{True}\mathrel{=}\Conid{False}} (\ensuremath{\mathrm{100}\mathbin{\%}}) and \ensuremath{\Varid{f}\;\Conid{False}} is either \ensuremath{\Conid{True}} (\ensuremath{\mathrm{95}\mathbin{\%}}) or \ensuremath{\Conid{False}} (\ensuremath{\mathrm{5}\mathbin{\%}})
--- an instance of \emph{faulty negation}.
It is easy to represent \ensuremath{\Varid{f}} in the form of a matrix \ensuremath{\Conid{M}},
\begin{eqnarray}
\ensuremath{\Conid{M}} =  ~~~~
\begin{matrix}
{\begin{matrix}
~~~~
& 
\ensuremath{\Conid{False}}
& ~
\ensuremath{\Conid{True}} 
\end{matrix}
} \\
{\begin{matrix}
\ensuremath{\Conid{False}} 
\\
\\
\ensuremath{\Conid{True}}
\end{matrix}
\begin{pmatrix}
\ensuremath{\mathrm{0.05}} && 1.00
\\
\\
\ensuremath{\mathrm{0.95}} &&0.00
\end{pmatrix}}
\end{matrix}
	\label{eq:130326f}
\end{eqnarray}
where the inputs spread across columns and the outputs across rows.
Because columns represent distributions, all figures in the same column should sum up to \ensuremath{\mathrm{1}}.

Matrices with this property will be referred to as \emph{column-stochastic}
(CS). The multiplication of two CS-matrices is a CS-matrix, as is the identity
matrix \ensuremath{\Varid{id}} (square, diagonal matrix with \ensuremath{\mathrm{1}}s in the diagonal) which is the
unit of such multiplication: \ensuremath{\Varid{\Conid{M}.id}\mathrel{=}\Conid{M}\mathrel{=}\Varid{id} \comp \Conid{M}}, where matrix multiplication
is denoted by an infix dot \ensuremath{( \comp )}.

We will write \ensuremath{\Conid{M}\mathbin{:}\Varid{n}\to \Varid{m}}, or draw the arrow  $\rarrow n M m$,
to indicate the \emph{type} of a CS-matrix \ensuremath{\Conid{M}}, meaning that it has
\ensuremath{\Varid{n}} columns and \ensuremath{\Varid{m}} rows. This view enables us to regard all CS-matrices
as morphisms of a category whose objects are matrix dimensions, each dimension
having its identity morphism \ensuremath{\Varid{id}}. 
If one extends such objects to arbitrary types
(with Cartesian product and disjoint union for addition and multiplication
of matrix dimensions), this category of matrices
turns out to represent the Kleisli category induced
by the (finite) distribution monad. In the example above, \ensuremath{\Varid{f}\mathbin{:}\Conid{Bool}\to \Conid{Dist}\;\Conid{Bool}}
is represented by a matrix of type \ensuremath{\Conid{M}\mathbin{:}\Conid{Bool}\to \Conid{Bool}} (\ref{eq:130326f})
on the Kleisli-category side.

Let notation \ensuremath{\mean{\Varid{f}}} mean the matrix which
represents probabilistic function \ensuremath{\Varid{f}} in such a CS-matrix category.
For \ensuremath{\Varid{f}} of type \ensuremath{\Conid{A}\to \Conid{Dist}\;\Conid{B}}, \ensuremath{\mean{\Varid{f}}} will be a matrix of type \ensuremath{\Conid{A}\to \Conid{B}},
that is, cell \ensuremath{\Varid{b}\;\mean{\Varid{f}}\;\Varid{a}} in the matrix
\footnote{Following the infix notation usually adopted for relations (which
are Boolean matrices), for instance \ensuremath{\Varid{y}\leq \Varid{x}}, we write \ensuremath{\Varid{y}\;\Conid{M}\;\Varid{x}} to denote the
contents of the cell in matrix \ensuremath{\Conid{M}} addressed by row \ensuremath{\Varid{y}} and column \ensuremath{\Varid{x}}. This
and other notational conventions of the linear algebra of programming are
explained in detail in \citep{Ol12b}.}
records the probability of
\ensuremath{\Varid{b}} in distribution \ensuremath{ \delta \mathrel{=}\Varid{f}\;\Varid{a}}. Then
probabilistic function (monadic) composition,
\begin{hscode}\SaveRestoreHook
\column{B}{@{}>{\hspre}l<{\hspost}@{}}%
\column{E}{@{}>{\hspre}l<{\hspost}@{}}%
\>[B]{}(\Varid{f} \kcomp \Varid{g})\;\Varid{a}\mathrel{=}\mathbf{do}\;\{\mskip1.5mu \Varid{b}\leftarrow \Varid{g}\;\Varid{a};\Varid{f}\;\Varid{b}\mskip1.5mu\}{}\<[E]%
\ColumnHook
\end{hscode}\resethooks
becomes matrix multiplication,
\begin{eqnarray}
\ensuremath{\mean{\Varid{f} \kcomp \Varid{g}}\mathrel{=}\mean{\Varid{f}} \comp \mean{\Varid{g}}}
\end{eqnarray}
and probabilistic function choice is given by
\begin{eqnarray}
	\ensuremath{\mean{{\Varid{f}}\choice{\Varid{p}}{\Varid{g}}}} &=& p \ensuremath{\mean{\Varid{f}}} + (1-p) \ensuremath{\mean{\Varid{g}}}
	\label{eq:111211c}
\end{eqnarray}
where \ensuremath{\mathbin{+}} denotes addition of two matrices of the same type and
\ensuremath{\Varid{p}\;\Conid{M}} denotes the multiplication of every cell in \ensuremath{\Conid{M}} by probability \ensuremath{\Varid{p}}.

Clearly, \ensuremath{\mean{\Varid{return}}\mathrel{=}\Varid{id}}. Any conventional function \ensuremath{\Varid{f}\mathbin{:}\Conid{A}\to \Conid{B}} can be turned into a 
``sharp'' probabilistic one through the composition \ensuremath{\Varid{return} \comp \Varid{f}} which,
represented as a CS-matrix, is the matrix \ensuremath{\Conid{M}\mathrel{=}\mean{\Varid{return} \comp \Varid{f}}}
such that \ensuremath{\Varid{b}\;\Conid{M}\;\Varid{a}\mathrel{=}\mathrm{1}} if \ensuremath{\Varid{b}\mathrel{=}\Varid{f}\;\Varid{a}} and is \ensuremath{\mathrm{0}} otherwise.\footnote{A probabilistic function
\ensuremath{\Varid{f}\mathbin{:}\Conid{A}\to \Conid{Dist}\;\Conid{B}} is said to be \emph{sharp} if, for all $a\in A$, \ensuremath{\Varid{f}\;\Varid{a}} is a Dirac
distribution. A Dirac distribution is one whose support is a singleton set, the
unique element of which is offered with \ensuremath{\mathrm{100}\mathbin{\%}} probability.}
We will write \ensuremath{\mean{\Varid{f}}} as shorthand for \ensuremath{\mean{\Varid{return} \comp \Varid{f}}}
and therefore will rely on fact \ensuremath{(\Varid{f}\;\Varid{a})\;\mean{\Varid{f}}\;\Varid{a}\mathrel{=}\mathrm{1}},
all other cells being \ensuremath{\mathrm{0}}.

The fact that sharp functions are representable by matrices
and that function composition corresponds to chaining the 
corresponding matrix arrows makes it easy to
picture probabilistic functional programs in the form of diagrams in the
matrix (Kleisli) category. Take, for instance, the
for-loop combinator given above,
\begin{hscode}\SaveRestoreHook
\column{B}{@{}>{\hspre}l<{\hspost}@{}}%
\column{E}{@{}>{\hspre}l<{\hspost}@{}}%
\>[B]{}\mathsf{for}\;\Varid{b}\;\Varid{i}\;\mathrm{0}\mathrel{=}\Varid{i}{}\<[E]%
\\
\>[B]{}\mathsf{for}\;\Varid{b}\;\Varid{i}\;(\Varid{n}\mathbin{+}\mathrm{1})\mathrel{=}\Varid{b}\;(\mathsf{for}\;\Varid{b}\;\Varid{i}\;\Varid{n}){}\<[E]%
\ColumnHook
\end{hscode}\resethooks
and re-write it as follows,
\begin{hscode}\SaveRestoreHook
\column{B}{@{}>{\hspre}l<{\hspost}@{}}%
\column{E}{@{}>{\hspre}l<{\hspost}@{}}%
\>[B]{}(\mathsf{for}\;\Varid{b}\;\Varid{i}) \comp \underline{\mathrm{0}}\mathrel{=}\underline{\Varid{i}}{}\<[E]%
\\
\>[B]{}((\mathsf{for}\;\Varid{b}\;\Varid{i}) \comp  \succ )\;\Varid{n}\mathrel{=}(\Varid{b} \comp (\mathsf{for}\;\Varid{b}\;\Varid{i}))\;\Varid{n}{}\<[E]%
\ColumnHook
\end{hscode}\resethooks
where \ensuremath{ \succ \;\Varid{n}\mathrel{=}\Varid{n}\mathbin{+}\mathrm{1}} and (recall) the under-bar notation denotes constant functions.
This is the same as writing the matrix equalities,
\begin{hscode}\SaveRestoreHook
\column{B}{@{}>{\hspre}l<{\hspost}@{}}%
\column{E}{@{}>{\hspre}l<{\hspost}@{}}%
\>[B]{}\mean{\mathsf{for}\;\Varid{b}\;\Varid{i}} \comp \mean{\underline{\mathrm{0}}}\mathrel{=}\mean{\underline{\Varid{i}}}{}\<[E]%
\\
\>[B]{}\mean{\mathsf{for}\;\Varid{b}\;\Varid{i}} \comp \mean{ \succ }\mathrel{=}\mean{\Varid{b}} \comp \mean{\mathsf{for}\;\Varid{b}\;\Varid{i}}{}\<[E]%
\ColumnHook
\end{hscode}\resethooks
which can be reduced to a single equality,
\begin{eqnarray}
\ensuremath{\mean{\mathsf{for}\;\Varid{b}\;\Varid{i}} \comp [\mean{\underline{\mathrm{0}}}|\mean{ \succ }]\mathrel{=}[\mean{\underline{\Varid{i}}}|(\mean{\Varid{b}} \comp \mean{\mathsf{for}\;\Varid{b}\;\Varid{i}})]}
	\label{eq:130326g}
\end{eqnarray}
by resorting to the \ensuremath{[\Conid{M}|\Conid{N}]} combinator which glues two matrices \ensuremath{\Conid{M}\mathbin{:}\Conid{A}\to \Conid{C}} and \ensuremath{\Conid{N}\mathbin{:}\Conid{B}\to \Conid{C}} side-by-side, yielding \ensuremath{[\Conid{M}|\Conid{N}]\mathbin{:}\Conid{A}\mathbin{+}\Conid{B}\to \Conid{C}}. As explained by \cite{MO13c}, this combinator --- which corresponds to
the relational ``junc'' operator of \cite{BM97} ---  is a universal construction in any
category of matrices, therefore satisfying (among others) the fusion law
\begin{eqnarray}
	P\comp \meither M N &=& \meither{P\comp M}{P\comp N} \label{eq:meither:fusion}
\end{eqnarray}
and (for suitably typed matrices) the equality law,
\begin{eqnarray}
	\meither M N = \meither P Q & \equiv & M = P \land N = Q
	\label{eq:101221g}
\end{eqnarray}
both silently used in the derivation above.

Our matrix semantics for the \ensuremath{\mathsf{for}}-loop combinator can still be simplified in two ways:
first, the \ensuremath{\mean{\cdot }} parentheses in (\ref{eq:130326g})
can be dropped, since we may assume they are implicitly
surrounding functions everywhere:
\begin{hscode}\SaveRestoreHook
\column{B}{@{}>{\hspre}l<{\hspost}@{}}%
\column{E}{@{}>{\hspre}l<{\hspost}@{}}%
\>[B]{}(\mathsf{for}\;\Varid{b}\;\Varid{i}) \comp [\underline{\mathrm{0}}| \succ ]\mathrel{=}[\underline{\Varid{i}}|(\Varid{b} \comp (\mathsf{for}\;\Varid{b}\;\Varid{i}))]{}\<[E]%
\ColumnHook
\end{hscode}\resethooks
Second, \ensuremath{[\underline{\Varid{i}}|(\Varid{b} \comp (\mathsf{for}\;\Varid{b}\;\Varid{i}))]} can be factored into composition
\ensuremath{[\underline{\Varid{i}}|\Varid{b}] \comp (\Varid{id}\oplus (\mathsf{for}\;\Varid{b}\;\Varid{i}))}, since absorption law
\begin{eqnarray}
	\meither M N \comp (P \oplus Q) &=& \meither {M \comp P}{N \comp Q}
	\label{eq:101221f}
\end{eqnarray}
holds, 
where \ensuremath{\cdot \oplus \cdot } is the matrix direct
sum (block) operation: \ensuremath{\Conid{M}\oplus \Conid{N}\mathrel{=}\mblock{\Conid{M}}{\mathrm{0}}{\mathrm{0}}{\Conid{N}}}.
Altogether, we get an equality of matrix compositions,
\begin{hscode}\SaveRestoreHook
\column{B}{@{}>{\hspre}l<{\hspost}@{}}%
\column{E}{@{}>{\hspre}l<{\hspost}@{}}%
\>[B]{}(\mathsf{for}\;\Varid{b}\;\Varid{i}) \comp [\underline{\mathrm{0}}| \succ ]\mathrel{=}[\underline{\Varid{i}}|\Varid{b}] \comp (\Varid{id}\oplus (\mathsf{for}\;\Varid{b}\;\Varid{i})){}\<[E]%
\ColumnHook
\end{hscode}\resethooks
which corresponds to the typed matrix diagram which follows,
\begin{eqnarray*}
\xymatrix@R=6ex{
	\N_0
		\ar@/^1pc/[rr]^-{\conv{in}=\msplit{\conv{{\kons 0}}}{\conv{\succ}}}
		\ar[d]_{\ensuremath{\mathsf{for}\;\Varid{b}\;\Varid{i}}}
&
	\iso
&
	1 + \N_0
		\ar@/^1pc/[ll]^-{in=\meither{{\kons 0}}{\succ}}
		\ar[d]^{id \oplus \ensuremath{(\mathsf{for}\;\Varid{b}\;\Varid{i})}}
\\
	B
&
&
	1 + B
		\ar@/^1pc/[ll]^-{\meither{{\kons i}}{b}}
}
\end{eqnarray*}
where symbol $\iso$ indicates that function \ensuremath{\inN \mathrel{=}[\underline{\mathrm{0}}| \succ ]}
is a bijection, and therefore its converse \ensuremath{\conv{\inN }} is also a function.
By the \emph{converse} \ensuremath{\conv{\Conid{M}}} of a matrix \ensuremath{\Conid{M}} we mean its transpose, that
is, \ensuremath{\Varid{x}\;\conv{\Conid{M}}\;\Varid{y}\mathrel{=}\Varid{y}\;\Conid{M}\;\Varid{x}} holds: the effect is that of swapping rows
with columns. The diagram also uses the \emph{split} combinator \ensuremath{\msplit{\cdot }{\cdot }}
which is the converse dual of \ensuremath{[\cdot |\cdot ]}:
\begin{eqnarray}
	\ensuremath{\conv{[\Conid{M}|\Conid{N}]}}&=& \ensuremath{\msplit{\conv{\Conid{M}}}{\conv{\Conid{N}}}}
\end{eqnarray}

Why does this diagram matter? First, it can be recognized as an instance
of a \emph{catamorphism} diagram \citep{BM97}, here interpreted in the category
of CS-matrices rather than in that of total functions or binary relations
--- the \emph{qualitative} to \emph{quantitative} shift promised in the introduction
of the paper. In fact, because composition is closed for CS-matrices and
these include sharp functions, \ensuremath{\Varid{b}} and \ensuremath{\underline{\Varid{i}}} can vary inside the CS-matrix
space and the diagram will still make sense. For instance, the base case,
which is represented by constant function \ensuremath{\underline{\Varid{i}}\mathbin{:}\mathrm{1}\to \N_0} --- a column
vector --- corresponds to the Dirac distribution on \ensuremath{\Varid{i}}, which can be changed
to any other distribution.

Moreover, because \ensuremath{\inN } is a bijection, not only the diagram tells that \ensuremath{\mathsf{for}\;\Varid{b}\;\Varid{i}} is a solution to the equation
\begin{eqnarray*} 
	\ensuremath{\Varid{k} \comp \inN \mathrel{=}[\underline{\Varid{i}}|\Varid{b}] \comp (\Varid{id}\oplus \Varid{k})}
\end{eqnarray*}
but it turns out that this is the unique solution:\footnote{The argument
is the same as in \citep{BM97} just by replacing the powerset monad by the
distribution monad.}
\begin{eqnarray} 
	\ensuremath{\Varid{k}\mathrel{=}\mathsf{for}\;\Varid{b}\;\Varid{i}} & \equiv & \ensuremath{\Varid{k} \comp \inN \mathrel{=}[\underline{\Varid{i}}|\Varid{b}] \comp (\Varid{id}\oplus \Varid{k})}
	\label{eq:130321b}
\end{eqnarray}
This unique solution can be computed as the fixpoint in \ensuremath{\Varid{k}} of equation
\begin{eqnarray}
	\ensuremath{\Varid{k}\mathrel{=}\underline{\Varid{i}} \comp \conv{\underline{\mathrm{0}}}\mathbin{+}\Varid{b} \comp \Varid{k} \comp \conv{ \succ }}
	\label{eq:130321a}
\end{eqnarray}
which is obtained from (\ref{eq:130321b}) above by use of the so-called `divide-and-conquer' law:
\begin{eqnarray}
	\meither M N\comp \msplit P Q & = & M\comp P + N\comp Q
	\label{eq:090403c}
\end{eqnarray}

Equation (\ref{eq:130321a}) tells how the matrix \ensuremath{\Varid{k}\mathrel{=}\mathsf{for}\;\Varid{b}\;\Varid{i}} is recursively filled up: first the outer-product
\ensuremath{\underline{\Varid{i}} \comp \conv{\underline{\mathrm{0}}}}, that is, the everywhere-$0$
matrix apart from the $1$ in cell $(i,0)$), which is added to
	\ensuremath{\Varid{b} \comp \underline{\Varid{i}} \comp \conv{\underline{\mathrm{0}}} \comp \conv{ \succ }}, 
and so on. For sharp \ensuremath{\Varid{b}}, this is \ensuremath{\underline{(\Varid{b}\;\Varid{i})} \comp \conv{\mathrm{1}}},
the \ensuremath{\Varid{n}}-th entry being \ensuremath{\underline{({\Varid{b}}^{\Varid{n}}\;\Varid{i})} \comp \conv{\Varid{n}}}.
Note that each contribution of the fixpoint ascending chain is a matrix which ``fills an empty column'',
thus ensuring that no column ever adds up to more than $1$.

Equation (\ref{eq:130321a}) also serves to emulate the construction of the fixpoint using matrix algebra
packages such as, for instance, \matlab.
In this case we build finite approximations of the
fixpoint helped by the corresponding diagram approximation, for inputs
at most \ensuremath{\Varid{n}} and at most \ensuremath{\Varid{m}} possible outputs:
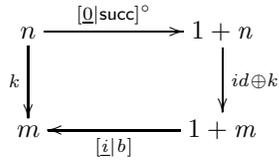
\begin{eqnarray*}
\xymatrix@C=5em{
	n
		\ar[r]^-{\ensuremath{\conv{[\underline{\mathrm{0}}| \succ ]}}}
		\ar[d]_{\ensuremath{\Varid{k}}}
&
	\ensuremath{\mathrm{1}\mathbin{+}\Varid{n}}
		\ar[d]^{\ensuremath{\Varid{id}\oplus \Varid{k}}}
\\
	m
&
	\ensuremath{\mathrm{1}\mathbin{+}\Varid{m}}
		\ar[l]^-{\ensuremath{[\underline{\Varid{i}}|\Varid{b}]}}
}
\end{eqnarray*}
As \matlab\ is not typed, tracing matrix dimensions
without the help of diagrams of this kind
would be a nightmare.

Let us see an example: suppose we want to emulate a fault in the \ensuremath{\Varid{odd}} function,
\ensuremath{\Varid{odd}\mathrel{=}(\mathrm{1}\mathbin{+}) \comp (\mathrm{2}\mathbin{*})}, in which \ensuremath{(\mathrm{2}\mathbin{*})\mathrel{=}\mathsf{for}\;(\mathrm{2}\mathbin{+})\;\mathrm{0}} is disturbed by the propagation
of the same fault of addition we have seen before:
\begin{hscode}\SaveRestoreHook
\column{B}{@{}>{\hspre}l<{\hspost}@{}}%
\column{E}{@{}>{\hspre}l<{\hspost}@{}}%
\>[B]{}\mathit{ftwice}_{\Varid{p}}\mathrel{=}\Varid{mfor}\;\mathit{fadd}_{\Varid{p}}\;\mathrm{2}\;\mathrm{0}{}\<[E]%
\ColumnHook
\end{hscode}\resethooks
For instance, \ensuremath{\mathit{ftwice}_{\mathrm{0.1}}\;\mathrm{4}} is the distribution
\begin{quote}\small
\begin{tabbing}\tt
~8~~65\char46{}6\char37{}\\
\tt ~6~~29\char46{}2\char37{}\\
\tt ~4~~~4\char46{}9\char37{}\\
\tt ~2~~~0\char46{}4\char37{}\\
\tt ~0~~~0\char46{}0\char37{}
\end{tabbing}
\end{quote}
In \matlab, we will first draw the corresponding diagram,
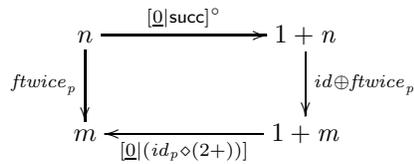
\begin{eqnarray*}
\xymatrix@C=6em{
	n
		\ar[r]^-{\ensuremath{\conv{[\underline{\mathrm{0}}| \succ ]}}}
		\ar[d]_{\ensuremath{\mathit{ftwice}_{\Varid{p}}}}
&
	\ensuremath{\mathrm{1}\mathbin{+}\Varid{n}}
		\ar[d]^{\ensuremath{\Varid{id}\oplus \mathit{ftwice}_{\Varid{p}}}}
\\
	m
&
	\ensuremath{\mathrm{1}\mathbin{+}\Varid{m}}
		\ar[l]^-{\ensuremath{[\underline{\mathrm{0}}|({\Varid{id}}\choice{\Varid{p}}{(\mathrm{2}\mathbin{+})})]}}
}
\end{eqnarray*}
parametric on probability \ensuremath{\Varid{p}} and the \ensuremath{\Varid{n}} and \ensuremath{\Varid{m}} dimensions, which nevertheless
have to be passed explicitly when encoding each arrow of the diagram as a \matlab\ matrix.
The probabilistic choice in the corresponding instance of
(\ref{eq:130321a}),
\begin{eqnarray}
	k & = & \ensuremath{\underline{\mathrm{0}} \comp \conv{\underline{\mathrm{0}}}\mathbin{+}({\Varid{id}}\choice{\Varid{p}}{(\mathrm{2}\mathbin{+})}) \comp \Varid{k} \comp \conv{ \succ }}
	\label{eq:130322a}
\end{eqnarray}
is captured by \matlab\ function
\begin{quote}\small
\begin{tabbing}\tt
~function~C~\char61{}~faddk\char40{}p\char44{}k\char44{}n\char44{}m\char41{}\\
\tt ~~~~M~\char61{}~eye\char40{}m\char44{}n\char41{}\char59{}\\
\tt ~~~~N~\char61{}~addk\char40{}k\char44{}n\char44{}m\char41{}\char59{}\\
\tt ~~~~C~\char61{}~choice\char40{}p\char44{}M\char44{}N\char41{}\char59{}\\
\tt ~end
\end{tabbing}
\end{quote}
(note the types, ie.\ dimensions \ensuremath{\Varid{n}} and \ensuremath{\Varid{m}}, passed as parameters)
where
\begin{quote}\small
\begin{tabbing}\tt
~function~C~\char61{}~choice\char40{}p\char44{}M\char44{}N\char41{}\\
\tt ~~~~~if~size\char40{}M\char41{}~\char126{}\char61{}~size\char40{}N\char41{}\\
\tt ~~~~~~~~~error\char40{}\char39{}Dimensions~must~agree\char39{}\char41{}\char59{}\\
\tt ~~~~~else\\
\tt ~~~~~~~~~C~\char61{}~p\char42{}M\char43{}\char40{}1\char45{}p\char41{}\char42{}N\\
\tt ~~~~end\\
\tt ~end
\end{tabbing}
\end{quote}
(note the need for explicit type error checking).
The right-hand side of the equation (\ref{eq:130322a}) is captured by
\begin{quote}\small
\begin{tabbing}\tt
~function~Y~\char61{}~twiceF\char40{}p\char44{}n\char44{}m\char44{}X\char41{}\\
\tt ~~~~if~size\char40{}X\char41{}~\char126{}\char61{}~\char91{}m~n\char93{}\\
\tt ~~~~~~~error\char40{}\char39{}Dimensions~must~agree\char39{}\char41{}\char59{}\\
\tt ~~~~else\\
\tt ~~~~~~~Y~\char61{}~zero\char40{}m\char41{}\char42{}zero\char40{}n\char41{}\char39{}~\char43{}\\
\tt ~~~~~~~~~~faddk\char40{}p\char44{}2\char44{}m\char44{}m\char41{}\char42{}X\char42{}succ\char40{}n\char44{}n\char41{}\char39{}\\
\tt ~~~~end\\
\tt ~end
\end{tabbing}
\end{quote}
For \ensuremath{\Varid{n},\Varid{m}\mathrel{=}\mathrm{5},\mathrm{8}} and \ensuremath{\Varid{p}\mathrel{=}\mathrm{0.1}}, the fixpoint of equation (\ref{eq:130322a}) is the matrix
\begin{quote}\small
\begin{tabbing}\tt
~1~~~~~~~0\char46{}1~~~~~0\char46{}01~~~~0\char46{}001~~~0\char46{}0001~~\\
\tt ~0~~~~~~~0~~~~~~~0~~~~~~~0~~~~~~~0\\
\tt ~0~~~~~~~0\char46{}9~~~~~0\char46{}18~~~~0\char46{}027~~~0\char46{}0036~~\\
\tt ~0~~~~~~~0~~~~~~~0~~~~~~~0~~~~~~~0\\
\tt ~0~~~~~~~0~~~~~~~0\char46{}81~~~~0\char46{}243~~~0\char46{}0486~~\\
\tt ~0~~~~~~~0~~~~~~~0~~~~~~~0~~~~~~~0\\
\tt ~0~~~~~~~0~~~~~~~0~~~~~~~0\char46{}729~~~0\char46{}2916~~\\
\tt ~0~~~~~~~0~~~~~~~0~~~~~~~0~~~~~~~0\\
\tt ~0~~~~~~~0~~~~~~~0~~~~~~~0~~~~~~~0\char46{}6561~~
\end{tabbing}
\end{quote}
whose leftmost column (resp.\ top row) corresponds to input (resp.\ output) \ensuremath{\mathrm{0}}.
The five columns of the matrix correspond to the
distributions output by the monadic \ensuremath{\mathit{ftwice}_{\mathrm{0.1}}\;\Varid{n}}, for \ensuremath{\Varid{n}\mathrel{=}\mathrm{0}\mathinner{\ldotp\ldotp}\mathrm{4}}.

So much for an illustration of the correspondence between monadic probabilistic programming
(in Haskell) and column stochastic matrix construction (in \matlab).
In the following section we will go back to analytical methods relying solely on
universal property (\ref{eq:130321b}) and its corollaries.

\section{Probabilistic mutual recursion in the LAoP}
\label{sec:130326b}
As we have seen above, mutual recursion arises from the \emph{pairing}
--- \emph{tupling}, in general \citep{HITT97} --- of two (sharp) functions \ensuremath{\Varid{f}} and \ensuremath{\Varid{g}},
defined by
\begin{quote}
	\ensuremath{(\Varid{f},\Varid{g})\;\Varid{x}\mathrel{=}(\Varid{f}\;\Varid{x},\Varid{g}\;\Varid{x})}
\end{quote}
where \ensuremath{(\Varid{f},\Varid{g})\mathbin{:}\Conid{A}\to \Conid{B} \times \Conid{C}} for \ensuremath{\Varid{f}\mathbin{:}\Conid{A}\to \Conid{B}} and \ensuremath{\Varid{g}\mathbin{:}\Conid{A}\to \Conid{C}}.
This tupling operator is known as \emph{split} in the functional
setting \citep{BM97} or as \emph{fork} in the relational one \citep{FBH97,Sc10}. \cite{Ma12ok}
shows that these operators generalize to the so-called Khatri-Rao product
\ensuremath{{\Conid{M}}\kr{\Conid{N}}} of two arbitrary matrices \ensuremath{\Conid{M}} and \ensuremath{\Conid{N}}, defined index-wise by
\begin{eqnarray}
	\ensuremath{(\Varid{b},\Varid{c})\;({\Conid{M}}\kr{\Conid{N}})\;\Varid{a}} &=& \ensuremath{(\Varid{b}\;\Conid{M}\;\Varid{a})} \times \ensuremath{(\Varid{c}\;\Conid{N}\;\Varid{a})}
	\label{eq:130129b}
\end{eqnarray}
Thus the Khatri-Rao product is a ``column-wise" version of the well-known
Kronecker product \ensuremath{\cdot \otimes \cdot }, defined by
\begin{eqnarray}
	\ensuremath{(\Varid{y},\Varid{x})\;(\Conid{M}\otimes \Conid{N})\;(\Varid{b},\Varid{a})} & = & \ensuremath{(\Varid{y}\;\Conid{M}\;\Varid{b})} \times \ensuremath{(\Varid{x}\;\Conid{N}\;\Varid{a})}
	\label{eq:120716a}
\end{eqnarray}
Khatri-Rao coincides with Kronecker for column vectors \ensuremath{\Varid{u}\mathbin{:}\mathrm{1}\to \Conid{B}}, \ensuremath{\Varid{v}\mathbin{:}\mathrm{1}\to \Conid{C}},
\begin{eqnarray}
	\ensuremath{{\Varid{u}}\kr{\Varid{v}}\mathrel{=}\Varid{u}\otimes \Varid{v}}
	\label{eq:130111a}
\end{eqnarray}
and commutes with matrix junc'ing via the \emph{exchange law} \citep{Ma12ok}:
\begin{eqnarray}
	\ensuremath{{[\Conid{M}|\Conid{N}]}\kr{[\Conid{P}|\Conid{Q}]}\mathrel{=}[({\Conid{M}}\kr{\Conid{P}})|({\Conid{N}}\kr{\Conid{Q}})]}
	\label{eq:110407a}
\end{eqnarray}
for suitably typed matrices \ensuremath{\Conid{M}}, \ensuremath{\Conid{N}}, \ensuremath{\Conid{P}} and \ensuremath{\Conid{Q}}.

For \emph{sharp} functions \ensuremath{\Varid{f}} and \ensuremath{\Varid{g}}, pairing is an universal construct
ensuring that any function \ensuremath{\Varid{k}} producing pairs is uniquely factored to the
left and to the right,
\begin{eqnarray}
	\ensuremath{\Varid{k}\mathrel{=}{\Varid{f}}\kr{\Varid{g}}} &\equiv & \ensuremath{\Varid{fst} \comp \Varid{k}\mathrel{=}\Varid{f}} \land \ensuremath{\Varid{snd} \comp \Varid{k}\mathrel{=}\Varid{g}}
	\label{eq:130323c}
\end{eqnarray}
where \ensuremath{\Varid{fst}\;(\Varid{b},\Varid{c})\mathrel{=}\Varid{b}} and \ensuremath{\Varid{snd}\;(\Varid{b},\Varid{c})\mathrel{=}\Varid{c}}. (Note how liberally we keep omitting
the \ensuremath{\mean{\cdot }} parentheses around the occurrence of functions inside matrix expressions.)

From (\ref{eq:130323c}) a number of useful corollaries arise, namely (keep
in mind that \ensuremath{\Varid{f}} and \ensuremath{\Varid{g}} should be sharp functions for the time being)
\emph{fusion},
\begin{eqnarray}
	\ensuremath{({\Varid{f}}\kr{\Varid{g}}) \comp \Varid{h}\mathrel{=}{(\Varid{f} \comp \Varid{h})}\kr{(\Varid{g} \comp \Varid{h})}}
	\label{eq:130323a}
\end{eqnarray}
\emph{reconstruction},\footnote{Cf.\ \emph{loss-less decomposition} \citep{Ol11}.}
\begin{eqnarray}
	\ensuremath{\Varid{k}\mathrel{=}{(\Varid{fst} \comp \Varid{k})}\kr{(\Varid{snd} \comp \Varid{k})}}
	\label{eq:130104b}
\end{eqnarray}
and pairwise \emph{equality}:
\begin{eqnarray}
	\ensuremath{{\Varid{k}}\kr{\Varid{h}}\mathrel{=}{\Varid{f}}\kr{\Varid{g}}} & \equiv & \ensuremath{\Varid{k}\mathrel{=}\Varid{f}} \land \ensuremath{\Varid{h}\mathrel{=}\Varid{g}}
	\label{eq:130323b}
\end{eqnarray}

This makes it easy to prove the mutual recursion law, below instantiated
to \ensuremath{\mathsf{for}}-loops, where \ensuremath{\mathsf{F}\;\Varid{f}} abbreviates \ensuremath{\Varid{id}\oplus \Varid{f}}:
\footnote{As is well-known, for sharp functions this law extends to other inductive types,
eg.\ lists, trees etc \citep{BM97,HITT97}.}
\begin{eqnarray*}
&&
	\ensuremath{{\Varid{f}}\kr{\Varid{g}}\mathrel{=}\mathsf{for}\;({\Varid{h}}\kr{\Varid{k}})\;(\Varid{i},\Varid{j})}
\just\equiv{ universal property (\ref{eq:130321a}) }
	\ensuremath{({\Varid{f}}\kr{\Varid{g}}) \comp \inN \mathrel{=}[\underline{(\Varid{i},\Varid{j})}|({\Varid{h}}\kr{\Varid{k}})] \comp \mathsf{F}\;({\Varid{f}}\kr{\Varid{g}})}
\just\equiv{ fusion (\ref{eq:130323a}) ; constant functions }
	\ensuremath{{(\Varid{f} \comp \inN )}\kr{(\Varid{g} \comp \inN )}\mathrel{=}[({\underline{\Varid{i}}}\kr{\underline{\Varid{j}}})|({\Varid{h}}\kr{\Varid{k}})] \comp \mathsf{F}\;({\Varid{f}}\kr{\Varid{g}})}
	\eqnnewpage
\just\equiv{ exchange law (\ref{eq:110407a}) }
	\ensuremath{{(\Varid{f} \comp \inN )}\kr{(\Varid{g} \comp \inN )}\mathrel{=}({[\underline{\Varid{i}}|\Varid{h}]}\kr{[\underline{\Varid{j}}|\Varid{k}]}) \comp \mathsf{F}\;({\Varid{f}}\kr{\Varid{g}})}
\just\equiv{ fusion (\ref{eq:130323a}) again }
	\ensuremath{{(\Varid{f} \comp \inN )}\kr{(\Varid{g} \comp \inN )}\mathrel{=}{([\underline{\Varid{i}}|\Varid{h}] \comp \mathsf{F}\;({\Varid{f}}\kr{\Varid{g}}))}\kr{([\underline{\Varid{j}}|\Varid{k}] \comp \mathsf{F}\;({\Varid{f}}\kr{\Varid{g}}))}}
\just\equiv{ equality (\ref{eq:130323b}) }
\begin{lcbr}
	\ensuremath{\Varid{f} \comp \inN \mathrel{=}[\underline{\Varid{i}}|\Varid{h}] \comp \mathsf{F}\;({\Varid{f}}\kr{\Varid{g}})}
\\
	\ensuremath{\Varid{g} \comp \inN \mathrel{=}[\underline{\Varid{j}}|\Varid{k}] \comp \mathsf{F}\;({\Varid{f}}\kr{\Varid{g}})}
\end{lcbr}
\end{eqnarray*}
Read in reverse direction, this reasoning explains how two recursive,
mutually dependent functions \ensuremath{\Varid{f}} and \ensuremath{\Varid{g}} (regarded as matrices)
combine with each other into one single function \ensuremath{{\Varid{f}}\kr{\Varid{g}}}, from which
one can extract both \ensuremath{\Varid{f}} and \ensuremath{\Varid{g}} by projecting according to the
\emph{cancellation} rule,
\begin{eqnarray}
	\ensuremath{\Varid{fst} \comp ({\Varid{f}}\kr{\Varid{g}})\mathrel{=}\Varid{f}} \wider\land \ensuremath{\Varid{snd} \comp ({\Varid{f}}\kr{\Varid{g}})\mathrel{=}\Varid{g}}
	\label{eq:130104a}
\end{eqnarray}
yet another corollary of (\ref{eq:130323c}).

The law just derived can be identified as the underpinning of the
(pointwise) derivations of \ensuremath{\Varid{fibl}} (resp.\ \ensuremath{\Varid{sql}}) from
\ensuremath{\Varid{fib}} (resp.\ \ensuremath{\Varid{sq}}) back to section \ref{sec:130320b}.
But note that \ensuremath{\Varid{f}} and \ensuremath{\Varid{g}} have been regarded as \emph{sharp}
functions thus far, and therefore what we have written is just a
rephrasing of what can be found already in the literature of
\emph{tupling}, see eg.\ references \citep{BM97,HITT97} among several others.

We are now interested in checking the probabilistic extension of (\ref{eq:130323c}).
Let two probabilistic functions \ensuremath{\Varid{f}} and \ensuremath{\Varid{g}} and their product \ensuremath{{\Varid{f}}\kr{\Varid{g}}}
be depicted as the CS-matrices of the following diagram:
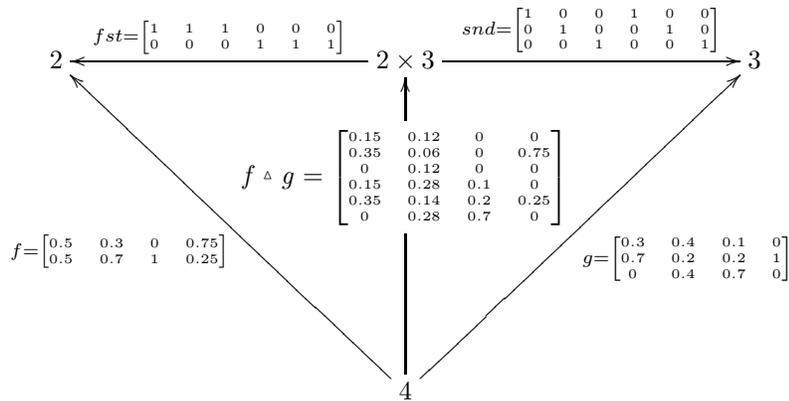
\begin{eqnarray*}
\xymatrix@C=25ex@R=25ex{
	2
&
	2\times 3
		\ar[l]_-{fst=\FST}
		\ar[r]^-{snd=\SND}
& 
	3
\\
&
	4
		\ar[u]|(.65)*+<1pt>+{\ensuremath{{\Varid{f}}\kr{\Varid{g}}} =\K}
		\ar[ru]_{g=\G}
		\ar[lu]^{f=\F}
}
\end{eqnarray*}
We can handle this in Haskell by running the
following monadic functions
\begin{hscode}\SaveRestoreHook
\column{B}{@{}>{\hspre}l<{\hspost}@{}}%
\column{E}{@{}>{\hspre}l<{\hspost}@{}}%
\>[B]{}({\Varid{f}}\kr{\Varid{g}})\;\Varid{a}\mathrel{=}\mathbf{do}\;\{\mskip1.5mu \Varid{b}\leftarrow \Varid{f}\;\Varid{a};\Varid{c}\leftarrow \Varid{g}\;\Varid{a};\Varid{return}\;(\Varid{b},\Varid{c})\mskip1.5mu\}{}\<[E]%
\\[\blanklineskip]%
\>[B]{}\Varid{mfst}\;\Varid{d}\mathrel{=}\mathbf{do}\;\{\mskip1.5mu (\Varid{b},\Varid{c})\leftarrow \Varid{d};\Varid{return}\;\Varid{b}\mskip1.5mu\}{}\<[E]%
\\[\blanklineskip]%
\>[B]{}\Varid{msnd}\;\Varid{d}\mathrel{=}\mathbf{do}\;\{\mskip1.5mu (\Varid{b},\Varid{c})\leftarrow \Varid{d};\Varid{return}\;\Varid{c}\mskip1.5mu\}{}\<[E]%
\ColumnHook
\end{hscode}\resethooks
inside the distribution monad \ensuremath{\Conid{Dist}},
thereby implementing the Khatri-Rao product and its projections.
For instance, \ensuremath{({\Varid{f}}\kr{\Varid{g}})\;\mathrm{2}} will yield
\begin{quote}\small
\begin{tabbing}\tt
~\char40{}2\char44{}1\char41{}~~28\char46{}0\char37{}\\
\tt ~\char40{}2\char44{}3\char41{}~~28\char46{}0\char37{}\\
\tt ~\char40{}2\char44{}2\char41{}~~14\char46{}0\char37{}\\
\tt ~\char40{}1\char44{}1\char41{}~~12\char46{}0\char37{}\\
\tt ~\char40{}1\char44{}3\char41{}~~12\char46{}0\char37{}\\
\tt ~\char40{}1\char44{}2\char41{}~~~6\char46{}0\char37{}
\end{tabbing}
\end{quote}
as in the second column of the corresponding matrix given above.
Moreover, both in Haskell and \matlab\ we can observe the
cancellations \ensuremath{\Varid{fst} \comp ({\Varid{f}}\kr{\Varid{g}})\mathrel{=}\Varid{f}} and \ensuremath{\Varid{snd} \comp ({\Varid{f}}\kr{\Varid{g}})\mathrel{=}\Varid{g}}.

However, \emph{reconstruction} (\ref{eq:130104b}) does not extend probabilistically.
This is because not every CS-matrix \ensuremath{\Varid{k}\mathbin{:}\Conid{A}\to \Conid{B} \times \Conid{C}}
outputting pairs is the Khatri-Rao product of two CS-matrices, as the
following counter-example shows: matrix
\begin{eqnarray*}
&&
	\ensuremath{\Varid{k}\mathbin{:}\mathrm{3}\to \mathrm{2} \times \mathrm{3}}
\\
&& k = \begin{bmatrix}
	0	& 0.4	& 0.2 \\
	0.2	& 0	& 0.17	\\
	0.2	& 0.1	& 0.13	\\
	0.6	& 0.4	& 0.2	\\
	0	& 0	& 0.17	\\
	0	& 0.1	& 0.13
\end{bmatrix}
\end{eqnarray*}
cannot be recovered from its projections, cf.\ the first column in:
\begin{eqnarray*}
	\ensuremath{{(\Varid{fst} \comp \Varid{k})}\kr{(\Varid{snd} \comp \Varid{k})}} =
\begin{bmatrix}
	0.24	& 0.4	& 0.2
\\	0.08	& 0	& 0.17
\\	0.08	& 0.1	& 0.13
\\	0.36	& 0.4	& 0.2
\\	0.12	& 0	& 0.17
\\	0.12	& 0.1	& 0.13
\end{bmatrix}
\end{eqnarray*}

This happens because probabilistic Khatri-Rao is a \emph{weak} product
--- the expected equivalence (\ref{eq:130323c}) is only an implication,
\begin{eqnarray}
	\ensuremath{\Varid{k}\mathrel{=}{\Varid{f}}\kr{\Varid{g}}} & \Rightarrow & \ensuremath{\Varid{fst} \comp \Varid{k}\mathrel{=}\Varid{f}} \land \ensuremath{\Varid{snd} \comp \Varid{k}\mathrel{=}\Varid{g}}
	\label{eq:121230a}
\end{eqnarray}
ensuring existence but not uniqueness. The proof of (\ref{eq:121230a}), which
is equivalent to cancellation (\ref{eq:130104a}) --- substitute \ensuremath{\Varid{k}} and simplify
--- can be found in appendix \ref{sec:130324a}. This proof relies on properties
(\ref{eq:130111a}) and (\ref{eq:110407a})
of the Khatri-Rao product.

Weak product (\ref{eq:121230a}) also grants pairwise equality (\ref{eq:130323b})
--- substitute \ensuremath{\Varid{k}} by \ensuremath{{\Varid{k}}\kr{\Varid{h}}} and simplify ---
but the converse substitution 
of \ensuremath{\Varid{f}} and \ensuremath{\Varid{g}}, in the $\Leftarrow$ direction,
leading to \emph{reconstruction} (\ref{eq:130104b}) is of course invalid.
In turn, this invalidates fusion (\ref{eq:130323a}) for
arbitrary probabilistic functions \ensuremath{\Varid{f}}, \ensuremath{\Varid{g}} and \ensuremath{\Varid{h}},
although the property will still hold in case \ensuremath{\Varid{h}} is sharp~\footnote{The same happens
with \emph{forks} in relation algebra \citep{BM97}.}, as
the straightforward proof in appendix \ref{sec:130324a} shows.

Altogether, the mutual recursion law will not hold in general for probabilistic
functions, as its calculation (above) relies on fusion (\ref{eq:130323a}).
This is consistent with what we have observed in section \ref{sec:130324b} concerning the
two versions of Fibonacci, \ensuremath{\Varid{mfib}} before the application of mutual recursion
and \ensuremath{\Varid{mfibl}} after, which differ substantially for inputs larger than \ensuremath{\mathrm{4}}.
However, the corresponding pair of probabilistic functions
of the other example --- \ensuremath{\Varid{msq}} and \ensuremath{\Varid{msql}} --- seemed to be the same
(ie.\ probabilistically indistinguishable), as much as
could be tested.

In the following section we explain the difference observed in the two
experiments by investigating sufficient conditions for the mutual recursion
law to hold for probabilistic functions (CS-matrices).

\section{Asymmetric Khatri-Rao product}
\label{sec:130326c}
To re-establish the equivalence in (\ref{eq:130323c}) given (\ref{eq:121230a})
we just have to find conditions for the converse implication
\begin{eqnarray*}
	\ensuremath{\Varid{k}\mathrel{=}{\Varid{f}}\kr{\Varid{g}}} & \Leftarrow & \ensuremath{\Varid{fst} \comp \Varid{k}\mathrel{=}\Varid{f}} \land \ensuremath{\Varid{snd} \comp \Varid{k}\mathrel{=}\Varid{g}}
\end{eqnarray*}
to hold, which is equivalent to (\ref{eq:130104b})
under the substitution or introduction of variables \ensuremath{\Varid{f}} and \ensuremath{\Varid{g}}.
For this we may seek inspiration in relation algebra, 
where one knows that if one of the projections of a binary
relation \ensuremath{\Conid{R}} outputting pairs is functional (ie., deterministic),
then \ensuremath{(\Varid{b},\Varid{c})\;\Conid{R}\;\Varid{a}\equiv \Varid{b}\;(\Varid{fst} \comp \Conid{R})\;\Varid{a}\;\land \;\Varid{c}\;(\Varid{snd} \comp \Conid{R})\;\Varid{a}} holds.
That is, by forking \ensuremath{\Varid{fst} \comp \Conid{R}} and \ensuremath{\Varid{snd} \comp \Conid{R}} one rebuilds \ensuremath{\Conid{R}}.

Back to probabilistic functions (ie.\ CS-matrices),
this suggests the 
conjecture:
\begin{quote}\em
If either \ensuremath{\Varid{fst} \comp \Varid{k}} or \ensuremath{\Varid{snd} \comp \Varid{k}} are sharp functions then (\ref{eq:130104b}) holds.
\end{quote}
Some intuitions first, before checking this. 
Let \ensuremath{\Varid{k}\mathbin{:}\Conid{A}\to \Conid{B} \times \Conid{C}} be a CS-matrix. The fact that \ensuremath{\Varid{f}\mathrel{=}\Varid{fst} \comp \Varid{k}\mathbin{:}\Conid{A}\to \Conid{B}} is sharp means that, for \ensuremath{\Varid{b}\mathrel{=}\Varid{f}\;\Varid{a}}, the corresponding \ensuremath{\Conid{C}}-block
in matrix \ensuremath{\Varid{k}} adds up to 1 and all the other entries in the \ensuremath{\Varid{a}}-column of
\ensuremath{\Varid{k}} are \ensuremath{\mathrm{0}}.  Projection \ensuremath{\Varid{snd} \comp \Varid{k}\mathbin{:}\Conid{A}\to \Conid{C}} yields such a block; \ensuremath{\conj{\Varid{fst} \comp \Varid{k}}{\Varid{snd} \comp \Varid{k}}} puts it back in place.

The proof of this conjecture, whereby (\ref{eq:130104b}) grants for free  the \emph{reflection law}
\begin{eqnarray}
	\ensuremath{\conj{\Varid{fst}}{\Varid{snd}}} &=& id
\end{eqnarray}
(take \ensuremath{\Varid{f},\Varid{g},\Varid{k}\mathbin{:=}\Varid{fst},\Varid{snd},\Varid{id}} and note that all functions involved are sharp),
will resort to the definition of (typed) matrix composition, for \ensuremath{\Conid{M}\mathbin{:}\Conid{B}\to \Conid{C}} and \ensuremath{\Conid{N}\mathbin{:}\Conid{A}\to \Conid{B}},
\begin{eqnarray}
	c(M \comp N)a &=& \ensuremath{\rcb\sum{\Varid{b}}{}{(\Varid{c}\;\Conid{M}\;\Varid{b}) \times (\Varid{b}\;\Conid{N}\;\Varid{a})}}
	\label{eq:120427a}
\end{eqnarray}
and to two rules 
which interface index-free and index-wise matrix notation, where $N$ is
an arbitrary matrix and $f$, $g$ are functional (ie.\ sharp) matrices:
\footnote{These rules
are derived by \cite{Ol12b} adopting the Eindhoven notation \citep{BM06,Mor12} for summations, eg.\
$\rcb\sum x R S$ where $R$ is the range (a predicate) which binds the dummy $x$ and $S$ is the summand.
$\rcb\sum x {} S$ corresponds to $R$ true for all $x$, the convention being omit $R$ in this case.}
\begin{eqnarray}
	y(f\comp N)x &\wider=& \rcb\sum z {\ensuremath{\Varid{y}\mathrel{=}\Varid{f}\;\Varid{z}}} {zNx}
	\label{eq:120428c}
\\
	y(\conv g \comp N\comp f)x &\wider=& \ensuremath{(\Varid{g}\;\Varid{y})\;\Conid{N}\;(\Varid{f}\;\Varid{x})}
	\label{eq:120428d}
\end{eqnarray}

Let us suppose \ensuremath{\Varid{fst} \comp \Varid{k}} in (\ref{eq:130104b}) is sharp. 
We denote by \ensuremath{\Varid{f}\mathbin{:}\Conid{A}\to \Conid{B}} the proper function which \ensuremath{\Varid{fst} \comp \Varid{k}} is,
by hypothesis. Thus \ensuremath{\Varid{f}\mathrel{=}\Varid{fst} \comp \Varid{k}}.
Regarded as a matrix, \ensuremath{\Varid{f}} is such that \ensuremath{\Varid{b}\;\Varid{f}\;\Varid{a}\mathrel{=}\mathrm{1}} if \ensuremath{\Varid{b}\mathrel{=}\Varid{f}\;\Varid{a}}, otherwise \ensuremath{\Varid{b}\;\Varid{f}\;\Varid{a}\mathrel{=}\mathrm{0}}.
It is easy to check that facts
\begin{eqnarray}
	&& \ensuremath{\rcb\sum{\Varid{c}}{}{(\Varid{f}\;\Varid{a},\Varid{c})\;\Varid{k}\;\Varid{a}}\mathrel{=}\mathrm{1}}
	\label{eq:130319a}
\\
	&& \ensuremath{\rcb\sum{(\Varid{b},\Varid{c})}{(\Varid{b} \not= \Varid{f}\;\Varid{a})}{((\Varid{b},\Varid{c})\;\Varid{k}\;\Varid{a})}\mathrel{=}\mathrm{0}}
	\label{eq:130319c}
\end{eqnarray}
hold --- see below.
Define \ensuremath{\Varid{m}\mathrel{=}\conj{\Varid{fst} \comp \Varid{k}}{\Varid{snd} \comp \Varid{k}}},
that is,
\begin{eqnarray*}
	\ensuremath{(\Varid{b},\Varid{c})\;\Varid{m}\;\Varid{a}} = \ensuremath{(\Varid{b}\;(\Varid{fst} \comp \Varid{k})\;\Varid{a}) \times (\Varid{c}\;(\Varid{snd} \comp \Varid{k})\;\Varid{a})}
\end{eqnarray*}
the same as
\begin{eqnarray}
	\ensuremath{(\Varid{b},\Varid{c})\;\Varid{m}\;\Varid{a}} = \ensuremath{(\Varid{b}\;\Varid{f}\;\Varid{a}) \times \rcb\sum{\Varid{b'}}{}{(\Varid{b'},\Varid{c})\;\Varid{k}\;\Varid{a}}}
	\label{eq:130319b}
\end{eqnarray}
since \ensuremath{\Varid{f}\mathrel{=}\Varid{fst} \comp \Varid{k}} an \ensuremath{\Varid{snd}} is sharp (\ref{eq:120428c}).
Our aim is to prove that \ensuremath{\Varid{m}\mathrel{=}\Varid{k}}.

\paragraph{Case \ensuremath{\Varid{b} \not= \Varid{f}\;\Varid{a}}:} In this case \ensuremath{\Varid{b}\;\Varid{f}\;\Varid{a}\mathrel{=}\mathrm{0}} and (\ref{eq:130319b})
yields \ensuremath{(\Varid{b},\Varid{c})\;\Varid{m}\;\Varid{a}\mathrel{=}\mathrm{0}}. From (\ref{eq:130319c}) we also get 
\ensuremath{(\Varid{b},\Varid{c})\;\Varid{k}\;\Varid{a}\mathrel{=}\mathrm{0}} and so \ensuremath{\Varid{m}\mathrel{=}\Varid{k}} for this case.

\paragraph{Case \ensuremath{\Varid{b}\mathrel{=}\Varid{f}\;\Varid{a}}:} we have
\begin{eqnarray*}
&&
	\ensuremath{(\Varid{f}\;\Varid{a},\Varid{c})\;\Varid{m}\;\Varid{a}}
\just={ (\ref{eq:130319b}) ; \ensuremath{(\Varid{b}\;\Varid{f}\;\Varid{a})\mathrel{=}\mathrm{1}} for \ensuremath{\Varid{b}\mathrel{=}\Varid{f}\;\Varid{a}}}
	\ensuremath{\rcb\sum{\Varid{b'}}{}{(\Varid{b'},\Varid{c})\;\Varid{k}\;\Varid{a}}}
\just={ \ensuremath{\Varid{b'}\mathrel{=}\Varid{f}\;\Varid{a}\; \mathbin\vee \;\Varid{b'} \not= \Varid{f}\;\Varid{a}}}
	\ensuremath{\rcb\sum{\Varid{b'}}{\Varid{b'}\mathrel{=}\Varid{f}\;\Varid{a}\; \mathbin\vee \;\Varid{b'} \not= \Varid{f}\;\Varid{a}}{(\Varid{b'},\Varid{c})\;\Varid{k}\;\Varid{a}}}
\just={ split summation ; one-point over \ensuremath{\Varid{b'}\mathrel{=}\Varid{f}\;\Varid{a}}}
	\ensuremath{((\Varid{f}\;\Varid{a},\Varid{c})\;\Varid{k}\;\Varid{a})} +
	\ensuremath{\rcb\sum{\Varid{b'}}{\Varid{b'} \not= \Varid{f}\;\Varid{a}}{(\Varid{b'},\Varid{c})\;\Varid{k}\;\Varid{a}}}
\just={ (\ref{eq:130319c}) }
	\ensuremath{(\Varid{f}\;\Varid{a},\Varid{c})\;\Varid{k}\;\Varid{a}}
\end{eqnarray*}
Thus \ensuremath{\Varid{m}} and \ensuremath{\Varid{k}} are extensionally the same for all cells addressed by \ensuremath{(\Varid{f}\;\Varid{a},\Varid{c})},
completing the proof.
\\ $\Box$

The proof assuming \ensuremath{\Varid{snd} \comp \Varid{k}} sharp instead of \ensuremath{\Varid{fst} \comp \Varid{k}} being so will be
essentially the same.
The remaining assumptions (\ref{eq:130319a}) and  (\ref{eq:130319c}) are easily proved in the appendix.

\section{Probabilistic mutual recursion resumed}
\label{sec:130326d}
Back to the case studies of section \ref{sec:130324b}, we
now capitalize on the result of the previous section granting that,
if one of the
projections of a probabilistic pair-valued function \ensuremath{\Varid{k}} is a sharp function,
then property (\ref{eq:130323c}) holds and all its corollaries.
\footnote{This includes, of course, the standard case in which both
\ensuremath{\Varid{f}} and \ensuremath{\Varid{g}} are sharp functions.}
This means that, under the same assumption, the mutual recursion law will hold too.

Put in other words, the probabilistic behaviour of a pair-valued recursive function,
for instance a \ensuremath{\mathsf{for}}-loop \ensuremath{\Varid{k}\mathrel{=}\mathsf{for}\;\Varid{b}\;\Varid{i}}, will be the same as the product \ensuremath{{\Varid{f}}\kr{\Varid{g}}} of its
mutually recursive projections \ensuremath{\Varid{f}} and \ensuremath{\Varid{g}}, provided either \ensuremath{\Varid{f}} is sharp or \ensuremath{\Varid{g}} is
sharp.

This enables us to spot a difference between the two examples of section
\ref{sec:130324b} just by looking at the corresponding call graphs:

\vskip 1em
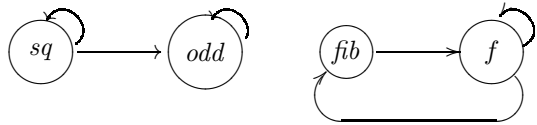
\begin{eqnarray*}
\SelectTips{cm}{}
\xymatrix @-1pc @C=7ex{
&
	*+++[o][F-]{ \mathit{sq} }
		\ar@(r,u)[]
		\ar[r]
&
	*+++[o][F-]{ \mathit{odd} }
		\ar@(r,u)[]
&
}
& \hskip -7em &
\xymatrix @-1pc @C=7ex{
&
	*++[o][F-]{ \mathit{fib} }
		\ar[r]
&
	*+++[o][F-]{ f }
		\ar@(r,u)[]
		\ar `dr_l[l] `_ur[l] ^{} [l]
&
}
\end{eqnarray*}
We see that \ensuremath{\Varid{sq}} depends on itself and on \ensuremath{\Varid{odd}} but \ensuremath{\Varid{odd}} only depends on itself.
Probabilistic \ensuremath{\Varid{msq}} was obtained from \ensuremath{\Varid{sq}} by injecting a fault in the addition
operation but this did not interfere with \ensuremath{\Varid{odd}}, which remained a sharp function.
Thus \ensuremath{\Varid{msql}} and \ensuremath{\Varid{msq}} exhibit the same probabilistic behaviour.

Comparatively, \ensuremath{\Varid{mfib}} was obtained from \ensuremath{\Varid{fib}} by injecting a similar fault but this
time the fault propagates to its derivative \ensuremath{\Varid{f}} and then back to \ensuremath{\Varid{mfib}}. Thus both
\ensuremath{\Varid{mfib}} and \ensuremath{\Varid{f}} are genuinely probabilistic and the derived linear version
\ensuremath{\Varid{mfibl}} is not granted to exhibit the same behaviour.

This can be confirmed by further querying our experiments in two ways.
First, we check that the \ensuremath{\Varid{odd}} projection of \ensuremath{\Varid{msql}} remains sharp
in spite of the probabilistic process it runs inside of:
we define \ensuremath{\Varid{msqlo}} as the same as \ensuremath{\Varid{msql}} but returning \ensuremath{\Varid{o}} instead of \ensuremath{\Varid{s}},
\begin{hscode}\SaveRestoreHook
\column{B}{@{}>{\hspre}l<{\hspost}@{}}%
\column{4}{@{}>{\hspre}l<{\hspost}@{}}%
\column{7}{@{}>{\hspre}l<{\hspost}@{}}%
\column{16}{@{}>{\hspre}l<{\hspost}@{}}%
\column{E}{@{}>{\hspre}l<{\hspost}@{}}%
\>[B]{}\Varid{msqlo}\;\Varid{n}\mathrel{=}{}\<[E]%
\\
\>[B]{}\hsindent{4}{}\<[4]%
\>[4]{}\mathbf{do}\;\{\mskip1.5mu (\Varid{s},\Varid{o})\leftarrow \Varid{mfor}\;\Varid{loop}\;(\mathrm{0},\mathrm{1})\;\Varid{n};\Varid{return}\;\Varid{o}\mskip1.5mu\}{}\<[E]%
\\
\>[4]{}\hsindent{3}{}\<[7]%
\>[7]{}\mathbf{where}\;\Varid{loop}\;(\Varid{s},\Varid{o})\mathrel{=}{}\<[E]%
\\
\>[7]{}\hsindent{9}{}\<[16]%
\>[16]{}\mathbf{do}\;\{\mskip1.5mu \Varid{z}\leftarrow \mathit{fadd}_{\mathrm{0.1}}\;\Varid{s}\;\Varid{o};\Varid{return}\;(\Varid{z},\Varid{o}\mathbin{+}\mathrm{2})\mskip1.5mu\}{}\<[E]%
\ColumnHook
\end{hscode}\resethooks
and run eg.\ \ensuremath{\Varid{msqlo}\;\mathrm{5}}, for instance
\begin{quote}\small
\begin{tabbing}\tt
~Main\char62{}~msqlo~5\\
\tt ~11~100\char46{}0\char37{}
\end{tabbing}
\end{quote}
to observe that it yields the Dirac distribution on \ensuremath{\mathrm{11}}, the fifth odd number,
 while its companion projection yields
\begin{quote}\small
\begin{tabbing}\tt
~Main\char62{}~msql~5\\
\tt ~25~~65\char46{}6\char37{}\\
\tt ~~9~~10\char46{}0\char37{}\\
\tt ~16~~~9\char46{}0\char37{}\\
\tt ~21~~~8\char46{}1\char37{}\\
\tt ~24~~~7\char46{}3\char37{}
\end{tabbing}
\end{quote}
Second, we disturb this situation by injecting another fault,
this time in the \ensuremath{\Varid{odd}} function itself,
\begin{hscode}\SaveRestoreHook
\column{B}{@{}>{\hspre}l<{\hspost}@{}}%
\column{E}{@{}>{\hspre}l<{\hspost}@{}}%
\>[B]{}\Varid{odd'}\;\mathrm{0}\mathrel{=}\Varid{return}\;\mathrm{1}{}\<[E]%
\\
\>[B]{}\Varid{odd'}\;(\Varid{n}\mathbin{+}\mathrm{1})\mathrel{=}\mathbf{do}\;\{\mskip1.5mu \Varid{x}\leftarrow \Varid{odd'}\;\Varid{n};\mathit{fadd}_{\mathrm{0.1}}\;\mathrm{2}\;\Varid{x}\mskip1.5mu\}{}\<[E]%
\ColumnHook
\end{hscode}\resethooks
and check that suitably adapted \ensuremath{\Varid{msq}},
mutually dependent on \ensuremath{\Varid{odd'}},
\begin{hscode}\SaveRestoreHook
\column{B}{@{}>{\hspre}l<{\hspost}@{}}%
\column{E}{@{}>{\hspre}l<{\hspost}@{}}%
\>[B]{}\Varid{msq'}\;\mathrm{0}\mathrel{=}\Varid{return}\;\mathrm{0}{}\<[E]%
\\
\>[B]{}\Varid{msq'}\;(\Varid{n}\mathbin{+}\mathrm{1})\mathrel{=}\mathbf{do}\;\{\mskip1.5mu \Varid{m}\leftarrow \Varid{msq'}\;\Varid{n};\Varid{x}\leftarrow \Varid{odd'}\;\Varid{n};\mathit{fadd}_{\mathrm{0.1}}\;\Varid{m}\;\Varid{x}\mskip1.5mu\}{}\<[E]%
\ColumnHook
\end{hscode}\resethooks
and its linear version,
\begin{hscode}\SaveRestoreHook
\column{B}{@{}>{\hspre}l<{\hspost}@{}}%
\column{4}{@{}>{\hspre}l<{\hspost}@{}}%
\column{8}{@{}>{\hspre}l<{\hspost}@{}}%
\column{12}{@{}>{\hspre}l<{\hspost}@{}}%
\column{E}{@{}>{\hspre}l<{\hspost}@{}}%
\>[B]{}\Varid{msql'}\;\Varid{n}\mathrel{=}{}\<[E]%
\\
\>[B]{}\hsindent{4}{}\<[4]%
\>[4]{}\mathbf{do}\;\{\mskip1.5mu (\Varid{s},\Varid{o})\leftarrow \Varid{mfor}\;\Varid{loop}\;(\mathrm{0},\mathrm{1})\;\Varid{n};\Varid{return}\;\Varid{s}\mskip1.5mu\}\;\mathbf{where}{}\<[E]%
\\
\>[4]{}\hsindent{4}{}\<[8]%
\>[8]{}\Varid{loop}\;(\Varid{s},\Varid{o})\mathrel{=}\mathbf{do}\;\{\mskip1.5mu {}\<[E]%
\\
\>[8]{}\hsindent{4}{}\<[12]%
\>[12]{}\Varid{z}\leftarrow \mathit{fadd}_{\mathrm{0.1}}\;\Varid{s}\;\Varid{o};\Varid{x}\leftarrow \mathit{fadd}_{\mathrm{0.1}}\;\mathrm{2}\;\Varid{o};{}\<[E]%
\\
\>[8]{}\hsindent{4}{}\<[12]%
\>[12]{}\Varid{return}\;(\Varid{z},\Varid{x})\mskip1.5mu\}{}\<[E]%
\ColumnHook
\end{hscode}\resethooks
now exhibit different probabilistic behaviours,
for instance,
\begin{quote}\small
\begin{tabular}{c|c|c}
\ensuremath{\Varid{n}} & \ensuremath{\Varid{msq'}\;\Varid{n}} & \ensuremath{\Varid{msql'}\;\Varid{n}}
\\\hline
	3
&
	\begin{minipage}{30em}
\begin{tabbing}\tt
~9~~59\char46{}0\char37{}\\
\tt ~7~~19\char46{}7\char37{}\\
\tt ~5~~10\char46{}3\char37{}\\
\tt ~8~~~6\char46{}6\char37{}\\
\tt ~6~~~2\char46{}2\char37{}\\
\tt ~3~~~1\char46{}9\char37{}\\
\tt ~4~~~0\char46{}2\char37{}\\
\tt ~1~~~0\char46{}1\char37{}\\
\tt ~2~~~0\char46{}0\char37{}
\end{tabbing}
	\end{minipage}
&
	\begin{minipage}{30em}
\begin{tabbing}\tt
~9~~65\char46{}6\char37{}\\
\tt ~5~~15\char46{}4\char37{}\\
\tt ~7~~~7\char46{}3\char37{}\\
\tt ~8~~~7\char46{}3\char37{}\\
\tt ~3~~~2\char46{}6\char37{}\\
\tt ~4~~~0\char46{}8\char37{}\\
\tt ~6~~~0\char46{}8\char37{}\\
\tt ~1~~~0\char46{}1\char37{}\\
\tt ~2~~~0\char46{}1\char37{}
\end{tabbing}
	\end{minipage}
\end{tabular}
\end{quote}
where linear scores better than mutually recursive, still.

\section{Generalizing to other fault propagation patterns}
\label{sec:130326e}
Besides mutual recursion, other fault propagation patterns
in functional programs arise from calculations in the LAoP.
These extend to other datatypes, as \ensuremath{\mathsf{for}}-loops generalize
to folds over lists, and more generally to
catamorphisms over other inductive data types \citep{BM97}.

Below we give examples of this generalization. The first example,
still dealing with \ensuremath{\mathsf{for}}-loops, shows that faults in
the base case propagate linearly through the choice operator
--- the law of \emph{base case fault distribution}:
\begin{eqnarray}
	\ensuremath{\mathsf{for}\;\Varid{f}\;({\Varid{a}}\choice{\Varid{p}}{\Varid{b}})} &=& \ensuremath{{(\mathsf{for}\;\Varid{f}\;\Varid{a})}\choice{\Varid{p}}{(\mathsf{for}\;\Varid{f}\;\Varid{b})}}  
	\label{eq:130220a}
\end{eqnarray}
The need for a generalization can be seen already in writing ``\ensuremath{{\Varid{a}}\choice{\Varid{p}}{\Varid{b}}}'',
an abuse of notation since the choice operator chooses between functions, not
arbitrary values.
Thus construct \ensuremath{\mathsf{for}\;\Varid{f}\;\Varid{i}} has to give room to \ensuremath{\mathopen{(\!|}[\Varid{h}|\Varid{f}]\mathclose{|\!)}}, where standard
catamorphism notation \citep{BM97} is adopted to give freedom to the base case to be
any probabilistic function \ensuremath{\Varid{h}} of its type. Thus (\ref{eq:130321b}) becomes,
for \ensuremath{\mathsf{F}\;\Varid{f}\mathrel{=}\Varid{id}\oplus \Varid{f}},
\begin{eqnarray} 
	\ensuremath{\Varid{k}\mathrel{=}\mathopen{(\!|}[\Varid{h}|\Varid{f}]\mathclose{|\!)}} & \equiv & \ensuremath{\Varid{k} \comp \inN \mathrel{=}[\Varid{h}|\Varid{f}] \comp (\mathsf{F}\;\Varid{k})}
	\label{eq:130325a}
\end{eqnarray}
Clearly,
\begin{eqnarray}
	\ensuremath{\mathsf{for}\;\Varid{f}\;\Varid{a}} &=& \ensuremath{\mathopen{(\!|}[\underline{\Varid{a}}|\Varid{f}]\mathclose{|\!)}}
	\label{eq:130325b}
\end{eqnarray}
holds. In (\ref{eq:130220a}), abbreviation \ensuremath{\mathsf{for}\;\Varid{f}\;({\Varid{a}}\choice{\Varid{p}}{\Varid{b}})}
replacing \ensuremath{\mathopen{(\!|}[({\underline{\Varid{a}}}\choice{\Varid{p}}{\underline{\Varid{b}}})|\Varid{f}]\mathclose{|\!)}}
is welcome as it enhances readability.

The proof of (\ref{eq:130220a}) is given in appendix \ref{sec:130324a}.
It relies on properties of probabilistic choice
already given by \cite{Ol12a}, namely \emph{choice-fusion}
\begin{eqnarray}
	\ensuremath{({\Varid{f}}\choice{\Varid{p}}{\Varid{g}}) \comp \Varid{h}} &=& \ensuremath{{(\Varid{f} \comp \Varid{h})}\choice{\Varid{p}}{(\Varid{g} \comp \Varid{h})}}
	\label{eq:111216b}
\\
	\ensuremath{\Varid{h} \comp ({\Varid{f}}\choice{\Varid{p}}{\Varid{g}})} &=& \ensuremath{{(\Varid{h} \comp \Varid{f})}\choice{\Varid{p}}{(\Varid{h} \comp \Varid{f})}}
	\label{eq:111219b}
\end{eqnarray}
and the \emph{exchange law}:
\begin{eqnarray}
	\ensuremath{{[\Varid{f}|\Varid{g}]}\choice{\Varid{p}}{[\Varid{h}|\Varid{k}]}} &=& \ensuremath{[({\Varid{f}}\choice{\Varid{p}}{\Varid{h}})|({\Varid{g}}\choice{\Varid{p}}{\Varid{k}})]}
	\label{eq:111219a}
\end{eqnarray}

Other interesting patterns of fault propagation arise in \emph{pipelining},
that is, compositions of probabilistic functions \ensuremath{\Varid{k}\mathrel{=}\Varid{f} \comp \Varid{g}} whereby one is able to
obtain the \emph{fault of the whole} (probabilistic \ensuremath{\Varid{k}}) expressed in terms of the
\emph{faults of its parts} (probabilistic \ensuremath{\Varid{f}} and \ensuremath{\Varid{g}}) by ``fault fusion''.

The example of fault fusion given below involves \emph{sequences} rather
than natural numbers, which means evolving from the \ensuremath{\mathsf{for}} combinator to the
corresponding combinator at sequence processing level \footnote{Both are
instances of the generic \emph{catamorphism} construct, as already mentioned.},
\begin{eqnarray} 
	\ensuremath{\Varid{k}\mathrel{=}\mathsf{fold}\;\Varid{f}\;\Varid{d}} & \equiv & \ensuremath{\Varid{k} \comp \inN \mathrel{=}[\Varid{d}|\Varid{f}] \comp (\mathsf{F}\;\Varid{k})}
	\label{eq:130325c}
\end{eqnarray}
where \ensuremath{\mathsf{F}\;\Varid{k}\mathrel{=}\Varid{id}\oplus (\Varid{id}\otimes \Varid{k})} and
\ensuremath{\inN \mathrel{=}[{\mathsf{nil}}|{\mathsf{cons}}]} is the initial algebra of sequences,
for (in Haskell notation) \ensuremath{{\mathsf{nil}}\;\anonymous \mathrel{=}[\mskip1.5mu \mskip1.5mu]} and \ensuremath{{\mathsf{cons}}\;(\Varid{h},\Varid{t})\mathrel{=}\Varid{h}\mathbin{:}\Varid{t}}.
Besides the direct sum (\ensuremath{\Varid{id}\oplus \cdot }) splitting base from recursive case,
as with \ensuremath{\mathsf{for}}, recursive pattern \ensuremath{\mathsf{F}\;\Varid{k}} involves the Kronecker product \ensuremath{\Varid{id}\otimes \Varid{k}}
which delivers to \ensuremath{\Varid{f}} the head of the current sequence and the
outcome of the recursive call \ensuremath{\Varid{k}}. The base case is captured by vector \ensuremath{\Varid{d}},
a distribution. For sharp functions, \ensuremath{\mathsf{fold}\;\Varid{f}\;\underline{\Varid{u}}} means the same
as \ensuremath{\mathsf{foldr}\;(\Varid{curry}\;\Varid{f})\;\Varid{u}} in standard Haskell. (This difference is not
a very significant one, as we shall see in the examples below.)
Substitution of \ensuremath{\Varid{k}} will yield a closed formula for probabilistic \ensuremath{\mathsf{fold}}
(cancellation property):
\begin{eqnarray}
	\ensuremath{\mathsf{fold}\;\Varid{f}\;\Varid{d}} &=& \ensuremath{[\Varid{d}|\Varid{f} \comp (\Varid{id}\otimes (\mathsf{fold}\;\Varid{f}\;\Varid{d}))] \comp \msplit{\conv{{\mathsf{nil}}}}{\conv{{\mathsf{cons}}}}}
	\nonumber
\just\equiv{ divide-and-conquer (\ref{eq:090403c}) }
	\ensuremath{\Varid{d} \comp \conv{{\mathsf{nil}}}\mathbin{+}\Varid{f} \comp (\Varid{id}\otimes (\mathsf{fold}\;\Varid{f}\;\Varid{d})) \comp \conv{{\mathsf{cons}}}}
	\label{eq:130325e}
\end{eqnarray}

As examples, consider \ensuremath{\Varid{count}\mathrel{=}\mathsf{fold}\;( \succ  \comp \Varid{snd})\;\underline{\mathrm{0}}}, the function that
counts how many items can be found in the input sequence,
and \ensuremath{\Varid{cat}\mathrel{=}\mathsf{fold}\;{\mathsf{cons}}\;{\mathsf{nil}}}, that which
copies the input sequence to the output (thus \ensuremath{\Varid{cat}\mathrel{=}\Varid{id}}).
Suppose there is some risk that \ensuremath{\Varid{cat}} might fail passing items from
input to output, with probability \ensuremath{\Varid{p}}, as captured by
\begin{hscode}\SaveRestoreHook
\column{B}{@{}>{\hspre}l<{\hspost}@{}}%
\column{E}{@{}>{\hspre}l<{\hspost}@{}}%
\>[B]{}\mathit{fcat}_{\Varid{p}}\mathrel{=}\mathsf{fold}\;({\Varid{lose}}\choice{\Varid{p}}{\Varid{send}})\;{\mathsf{nil}}{}\<[E]%
\ColumnHook
\end{hscode}\resethooks
where \ensuremath{\Varid{lose}\mathrel{=}\Varid{snd}} and \ensuremath{\Varid{send}\mathrel{=}{\mathsf{cons}}}.
\def\x{%
\begin{hscode}\SaveRestoreHook
\column{B}{@{}>{\hspre}l<{\hspost}@{}}%
\column{E}{@{}>{\hspre}l<{\hspost}@{}}%
\>[B]{}\underline{\cdot }\mathrel{=}\Varid{return}{}\<[E]%
\\
\>[B]{}{\mathsf{nil}}\mathrel{=}\underline{[\mskip1.5mu \mskip1.5mu]}{}\<[E]%
\\
\>[B]{}\Varid{lose}\;(\Varid{a},\Varid{b})\mathrel{=}\Varid{b}{}\<[E]%
\\
\>[B]{}\Varid{send}\;(\Varid{a},\Varid{b})\mathrel{=}\Varid{a}\mathbin{:}\Varid{b}{}\<[E]%
\ColumnHook
\end{hscode}\resethooks
}
For instance, for \ensuremath{\Varid{p}\mathrel{=}\mathrm{0.1}}, distribution \ensuremath{\mathit{fcat}_{\mathrm{0.1}}\;\text{\tt \char34 abc\char34}} will range from perfect copy (\ensuremath{\mathrm{72.9}\mathbin{\%}})
to complete loss (\ensuremath{\mathrm{0.1}\mathbin{\%}}):
\begin{quote}\small
\begin{tabbing}\tt
~\char34{}abc\char34{}~~72\char46{}9\char37{}\\
\tt ~~\char34{}ab\char34{}~~~8\char46{}1\char37{}\\
\tt ~~\char34{}ac\char34{}~~~8\char46{}1\char37{}\\
\tt ~~\char34{}bc\char34{}~~~8\char46{}1\char37{}\\
\tt ~~~\char34{}a\char34{}~~~0\char46{}9\char37{}\\
\tt ~~~\char34{}b\char34{}~~~0\char46{}9\char37{}\\
\tt ~~~\char34{}c\char34{}~~~0\char46{}9\char37{}\\
\tt ~~~~\char34{}\char34{}~~~0\char46{}1\char37{}
\end{tabbing}
\end{quote}
Now suppose that \ensuremath{\Varid{count}} too may be faulty in the sense of skipping elements
with probability \ensuremath{\Varid{q}}:
\begin{hscode}\SaveRestoreHook
\column{B}{@{}>{\hspre}l<{\hspost}@{}}%
\column{E}{@{}>{\hspre}l<{\hspost}@{}}%
\>[B]{}\mathit{fcount}_{\Varid{q}}\mathrel{=}\mathsf{fold}\;(({\Varid{id}}\choice{\Varid{q}}{ \succ }) \comp \Varid{snd})\;\underline{\mathrm{0}}{}\<[E]%
\ColumnHook
\end{hscode}\resethooks
For instance, for \ensuremath{\Varid{q}\mathrel{=}\mathrm{0.15}}, distribution \ensuremath{\mathit{fcount}_{\mathrm{0.15}}\;\text{\tt \char34 abc\char34}} will be:
\begin{quote}\small
\begin{tabbing}\tt
~3~~61\char46{}4\char37{}\\
\tt ~2~~32\char46{}5\char37{}\\
\tt ~1~~~5\char46{}7\char37{}\\
\tt ~0~~~0\char46{}3\char37{}
\end{tabbing}
\end{quote}

What can we tell about the risk of faults in the pipeline \ensuremath{\mathit{fcount}_{\Varid{q}} \comp \mathit{fcat}_{\cdot }}?
We could try specific runs, eg.\ \ensuremath{(\mathit{fcount}_{\Varid{q}} \comp \mathit{fcat}_{\Varid{p}})\;\text{\tt \char34 abc\char34}} yielding distribution
\begin{quote}\small
\begin{tabbing}\tt
~3~~44\char46{}8\char37{}\\
\tt ~2~~41\char46{}3\char37{}\\
\tt ~1~~12\char46{}7\char37{}\\
\tt ~0~~~1\char46{}3\char37{}
\end{tabbing}
\end{quote}
whose figures combine, \emph{in some way}, those given earlier for the individual
runs.

What we would like to know is the \emph{general} formula which combines
such figures and expresses the overall risk of failure.
For this we resort to the \emph{fusion law} which emerges from 
(\ref{eq:130325c}) in the standard way \citep{BM97} and also in the
probabilistic setting:
\begin{eqnarray}
	\ensuremath{\Varid{k} \comp (\mathsf{fold}\;\Varid{g}\;\Varid{e})\mathrel{=}\mathsf{fold}\;\Varid{f}\;\Varid{d}} & \implied & \ensuremath{\Varid{k} \comp [\Varid{e}|\Varid{g}]\mathrel{=}[\Varid{d}|\Varid{f}] \comp (\mathsf{F}\;\Varid{k})}
	\label{eq:130325d}
\end{eqnarray}
In our case, this enables us to solve the equation \ensuremath{\mathit{fcount}_{\Varid{q}} \comp \mathit{fcat}_{\Varid{p}}\mathrel{=}\mathsf{fold}\;\Varid{x}\;\Varid{y}}
for unknowns \ensuremath{\Varid{x}} and \ensuremath{\Varid{y}}:
\begin{eqnarray*}
&&
	\ensuremath{\mathit{fcount}_{\Varid{q}} \comp \mathit{fcat}_{\Varid{p}}\mathrel{=}\mathsf{fold}\;\Varid{x}\;\Varid{y}}
\just\implied{ \ensuremath{\mathsf{fold}} fusion (\ref{eq:130325d}) ; definition of \ensuremath{\mathit{fcat}_{\Varid{p}}} }
	\ensuremath{\mathit{fcount}_{\Varid{q}} \comp [{\mathsf{nil}}|({\Varid{lose}}\choice{\Varid{p}}{\Varid{send}})]} = \ensuremath{[\Varid{x}|\Varid{y}] \comp (\mathsf{F}\;\mathit{fcount}_{\Varid{q}})}
\just\equiv{ (\ref{eq:meither:fusion}) ; definition of \ensuremath{\mathsf{F}}; (\ref{eq:101221f}) ; (\ref{eq:101221g}) }
\begin{lcbr}
	\ensuremath{\mathit{fcount}_{\Varid{q}} \comp {\mathsf{nil}}\mathrel{=}\Varid{x}}
\\
	\ensuremath{\mathit{fcount}_{\Varid{q}} \comp ({\Varid{lose}}\choice{\Varid{p}}{\Varid{send}})\mathrel{=}\Varid{y} \comp (\Varid{id}\otimes \mathit{fcount}_{\Varid{q}})}
\end{lcbr}
\just\equiv{ \ensuremath{\mathit{fcount}_{\Varid{q}} \comp {\mathsf{nil}}\mathrel{=}\underline{\mathrm{0}}}}
\begin{lcbr}
	\ensuremath{\Varid{x}\mathrel{=}\underline{\mathrm{0}}}
\\
	\ensuremath{\mathit{fcount}_{\Varid{q}} \comp ({\Varid{snd}}\choice{\Varid{p}}{{\mathsf{cons}}})\mathrel{=}\Varid{y} \comp (\Varid{id}\otimes \mathit{fcount}_{\Varid{q}})}
\end{lcbr}
\end{eqnarray*}
Second, we solve the second equality just above for \ensuremath{\Varid{y}}:
\begin{eqnarray*}
&&
	\ensuremath{\mathit{fcount}_{\Varid{q}} \comp ({\Varid{snd}}\choice{\Varid{p}}{{\mathsf{cons}}})\mathrel{=}\Varid{y} \comp (\Varid{id}\otimes \mathit{fcount}_{\Varid{q}})}
\just\equiv{ choice fusion (\ref{eq:111219b}) }
	\ensuremath{{(\mathit{fcount}_{\Varid{q}} \comp \Varid{snd})}\choice{\Varid{p}}{(\mathit{fcount}_{\Varid{q}} \comp {\mathsf{cons}})}\mathrel{=}\Varid{y} \comp (\Varid{id}\otimes \mathit{fcount}_{\Varid{q}})}
\just\equiv{ unfolding \ensuremath{\mathit{fcount}_{\Varid{q}} \comp {\mathsf{cons}}}}
	\ensuremath{{(\mathit{fcount}_{\Varid{q}} \comp \Varid{snd})}\choice{\Varid{p}}{(({\Varid{id}}\choice{\Varid{q}}{ \succ }) \comp \Varid{snd} \comp (\Varid{id}\otimes \mathit{fcount}_{\Varid{q}}))}}
	\\ && = \ensuremath{\Varid{y} \comp (\Varid{id}\otimes \mathit{fcount}_{\Varid{q}})}
\just\equiv{ free theorem of \ensuremath{\Varid{snd}} }
	\ensuremath{{(\mathit{fcount}_{\Varid{q}} \comp \Varid{snd})}\choice{\Varid{p}}{(({\Varid{id}}\choice{\Varid{q}}{ \succ }) \comp \mathit{fcount}_{\Varid{q}} \comp \Varid{snd})}}
	\\ && = \ensuremath{\Varid{y} \comp (\Varid{id}\otimes \mathit{fcount}_{\Varid{q}})}
\just\equiv{ choice fusion (\ref{eq:111216b}) }
	\ensuremath{({\Varid{id}}\choice{\Varid{p}}{({\Varid{id}}\choice{\Varid{q}}{ \succ })}) \comp \mathit{fcount}_{\Varid{q}} \comp \Varid{snd}} = \ensuremath{\Varid{y} \comp (\Varid{id}\otimes \mathit{fcount}_{\Varid{q}})}
\just\equiv{ free theorem of \ensuremath{\Varid{snd}} again }
	\ensuremath{({\Varid{id}}\choice{\Varid{p}}{({\Varid{id}}\choice{\Varid{q}}{ \succ })}) \comp \Varid{snd} \comp (\Varid{id}\otimes \mathit{fcount}_{\Varid{q}})} = \ensuremath{\Varid{y} \comp (\Varid{id}\otimes \mathit{fcount}_{\Varid{q}})}
\just\implied{Leibniz (\ensuremath{\Varid{id}\otimes \mathit{fcount}_{\cdot }} cancelled from both sides)}
	\ensuremath{\Varid{y}\mathrel{=}({\Varid{id}}\choice{\Varid{p}}{({\Varid{id}}\choice{\Varid{q}}{ \succ })}) \comp \Varid{snd}}
\end{eqnarray*}
Summing up, we have been able to consolidate the
risk of the pipeline \ensuremath{\mathit{fcount}_{\Varid{q}} \comp \mathit{fcat}_{\Varid{p}}},
obtaining the overall behavior
\begin{hscode}\SaveRestoreHook
\column{B}{@{}>{\hspre}l<{\hspost}@{}}%
\column{6}{@{}>{\hspre}l<{\hspost}@{}}%
\column{10}{@{}>{\hspre}l<{\hspost}@{}}%
\column{E}{@{}>{\hspre}l<{\hspost}@{}}%
\>[B]{}\mathit{fcount}_{\Varid{q}} \comp \mathit{fcat}_{\Varid{p}}\mathrel{=}{}\<[E]%
\\
\>[B]{}\hsindent{6}{}\<[6]%
\>[6]{}\mathsf{fold}\;\Varid{y}\;\underline{\mathrm{0}}\;\mathbf{where}{}\<[E]%
\\
\>[6]{}\hsindent{4}{}\<[10]%
\>[10]{}\Varid{y}\mathrel{=}((\Varid{p}\mathbin{+}\Varid{q}\mathbin{-}\Varid{pq})\;\Varid{id}\mathbin{+}(\mathrm{1}\mathbin{-}\Varid{p})\;(\mathrm{1}\mathbin{-}\Varid{q})\; \succ ) \comp \Varid{snd}{}\<[E]%
\ColumnHook
\end{hscode}\resethooks
in which the probabilistic definition of \ensuremath{\Varid{y}} combines the choices according
to (\ref{eq:111211c}).
It can be checked that this behaviour
(which corresponds to that of a even more risky \ensuremath{\mathit{fcount}_{\cdot }} reading from a
perfect \ensuremath{\Varid{cat}}) matches up with the distributions
obtained for the specific runs given earlier.

\section{Probabilistic ``banana-split''}
Our final result has to do with a program transformation technique known
as \emph{banana-split} \citep{BM97}. Suppose you want to compute the average
of a non-empty list of integers:
\begin{eqnarray}
	\ensuremath{\Varid{avg}\;\Varid{l}} &=&\frac{\ensuremath{\Varid{sum}\;\Varid{l}}}{\ensuremath{\Varid{count}\;\Varid{l}}}
	\label{eq:131011a}
\end{eqnarray}
Clearly, you need to visit the input list \ensuremath{\Varid{l}} twice, one for
computing the sum of all integers and the other for knowing
how many there are.
\emph{Banana-split} is known as a corollary of the mutual recursion
law which enables one to merge \emph{both} visits into a single
one by keeping both values (current sum and current count) in
a pair.

From the results of section \ref{sec:130326d} one cannot take \emph{banana-split}
for granted in presence of faults, as mutual-recursion does not hold in general.
Let us start with an example: we inject faults in (\ref{eq:131011a}) 
by defining
\begin{hscode}\SaveRestoreHook
\column{B}{@{}>{\hspre}l<{\hspost}@{}}%
\column{13}{@{}>{\hspre}l<{\hspost}@{}}%
\column{E}{@{}>{\hspre}l<{\hspost}@{}}%
\>[B]{}\mathit{favg}_{\Varid{p},\Varid{q}}\mathrel{=}{}\<[13]%
\>[13]{}{\mathit{fsum}_{\Varid{p}}}\kr{\mathit{fcount}_{\Varid{q}}}{}\<[E]%
\ColumnHook
\end{hscode}\resethooks
for \ensuremath{\mathit{fcount}_{\Varid{q}}} as before and
\begin{hscode}\SaveRestoreHook
\column{B}{@{}>{\hspre}l<{\hspost}@{}}%
\column{10}{@{}>{\hspre}l<{\hspost}@{}}%
\column{E}{@{}>{\hspre}l<{\hspost}@{}}%
\>[B]{}\mathit{fsum}_{\Varid{p}}{}\<[10]%
\>[10]{}\mathrel{=}\mathsf{fold}\;(\Varid{uncurry}\;\mathit{fadd}_{\Varid{p}})\;\underline{\mathrm{0}}{}\<[E]%
\ColumnHook
\end{hscode}\resethooks
a (faulty) list sum function.\footnote{We focus on computing the pair of values
of  (\ref{eq:131011a}), leaving aside the final division and the problem of the
divisions by zero which arise from faulty counting (to be handled by raising exceptions).}
For instance, we have the outcome:
\begin{quote}\small
\begin{tabbing}\tt
~Main\char62{}~favg~0\char46{}15~0\char46{}1~\char91{}2\char44{}3\char93{}\\
\tt ~\char40{}5\char44{}2\char41{}~~58\char46{}5\char37{}\\
\tt ~\char40{}5\char44{}1\char41{}~~13\char46{}0\char37{}\\
\tt ~\char40{}2\char44{}2\char41{}~~10\char46{}3\char37{}\\
\tt ~\char40{}3\char44{}2\char41{}~~10\char46{}3\char37{}\\
\tt ~\char40{}2\char44{}1\char41{}~~~2\char46{}3\char37{}\\
\tt ~\char40{}3\char44{}1\char41{}~~~2\char46{}3\char37{}\\
\tt ~\char40{}0\char44{}2\char41{}~~~1\char46{}8\char37{}\\
\tt ~\char40{}5\char44{}0\char41{}~~~0\char46{}7\char37{}\\
\tt ~\char40{}0\char44{}1\char41{}~~~0\char46{}4\char37{}\\
\tt ~\char40{}2\char44{}0\char41{}~~~0\char46{}1\char37{}\\
\tt ~\char40{}3\char44{}0\char41{}~~~0\char46{}1\char37{}\\
\tt ~\char40{}0\char44{}0\char41{}~~~0\char46{}0\char37{}
\end{tabbing}
\end{quote}
which will lead to the correct average $2.5=\frac 5 2$ with \ensuremath{\mathrm{58.5}\mathbin{\%}} probability,
the wrong average of \ensuremath{\mathrm{5}} with \ensuremath{\mathrm{13.0}\mathbin{\%}} probability and so on and so forth.

By application of \emph{banana split} (details below) we transform \ensuremath{\mathit{favg}_{\Varid{p},\Varid{q}}} into a single
fold on total/count pairs \ensuremath{(\Varid{t},\Varid{c})},
\begin{hscode}\SaveRestoreHook
\column{B}{@{}>{\hspre}l<{\hspost}@{}}%
\column{4}{@{}>{\hspre}l<{\hspost}@{}}%
\column{9}{@{}>{\hspre}l<{\hspost}@{}}%
\column{15}{@{}>{\hspre}l<{\hspost}@{}}%
\column{E}{@{}>{\hspre}l<{\hspost}@{}}%
\>[B]{}\mathit{favgbs}_{\Varid{p},\Varid{q}}\mathrel{=}{}\<[15]%
\>[15]{}\mathsf{fold}\;\Varid{body}\;\underline{(\mathrm{0},\mathrm{0})}\;\mathbf{where}{}\<[E]%
\\
\>[B]{}\hsindent{4}{}\<[4]%
\>[4]{}\Varid{body}\;(\Varid{a},(\Varid{t},\Varid{c}))\mathrel{=}\mathbf{do}\;\{\mskip1.5mu {}\<[E]%
\\
\>[4]{}\hsindent{5}{}\<[9]%
\>[9]{}\Varid{t'}\leftarrow \mathit{fadd}_{\Varid{p}}\;\Varid{a}\;\Varid{t};{}\<[E]%
\\
\>[4]{}\hsindent{5}{}\<[9]%
\>[9]{}\Varid{c'}\leftarrow ({\Varid{id}}\choice{\Varid{q}}{ \succ })\;\Varid{c};{}\<[E]%
\\
\>[4]{}\hsindent{5}{}\<[9]%
\>[9]{}\Varid{return}\;(\Varid{t'},\Varid{c'})\mskip1.5mu\}{}\<[E]%
\ColumnHook
\end{hscode}\resethooks
which happens to yield the same output for the same arguments.

Perhaps the run above is not a good choice after all for showing some possible discrepancy between the
two versions of the code, before and after \emph{banana split} --- one would
say. It turns out that further experiments won't succeed in finding a run
discriminating both solutions, as these will remain probabilistically
indistinguishable.

We show below that this happens because the \emph{banana split} program transformation law \emph{does}
hold probabilistically, independently of mutual recursion.
To give a single proof covering both for-loops and folds on lists, we generalize both 
(\ref{eq:130321b}) and  (\ref{eq:130325c}) to 
\begin{eqnarray} 
	\ensuremath{\Varid{k}\mathrel{=}\mathopen{(\!|}\Varid{f}\mathclose{|\!)}} & \equiv & \ensuremath{\Varid{k} \comp \inN \mathrel{=}\Varid{f} \comp (\mathsf{F}\;\Varid{k})}
	\label{eq:131014b}
\end{eqnarray}
where \ensuremath{\Varid{f}} is a suitably typed probabilistic function covering both the inductive
and the base cases of (\ref{eq:130321b},\ref{eq:130325c}), and the customary
\emph{banana} brackets \ensuremath{\mathopen{(\!|}\anonymous \mathclose{|\!)}} are used to denote such a generic fold, or
\emph{catamorphism}.\footnote{Functors \ensuremath{\mathsf{F}\;\Conid{X}\mathrel{=}\Varid{id}\oplus \Conid{X}} and \ensuremath{\mathsf{F}\;\Conid{X}\mathrel{=}\Varid{id}\oplus (\Varid{id}\otimes \Conid{X})} give us back \ensuremath{\mathsf{for}}-loops and list folds, respectively.}
Cancellation
\begin{eqnarray}
	\ensuremath{\mathopen{(\!|}\Varid{f}\mathclose{|\!)} \comp \inN \mathrel{=}\Varid{f} \comp \mathsf{F}\;\mathopen{(\!|}\Varid{f}\mathclose{|\!)}}
	\label{eq:131014a}
\end{eqnarray}
follows trivially from (\ref{eq:131014b}).

\begin{theorem}[Probabilistic `banana-split']
Transformation
\begin{eqnarray}
	\ensuremath{{\mathopen{(\!|}\Varid{f}\mathclose{|\!)}}\kr{\mathopen{(\!|}\Varid{g}\mathclose{|\!)}}} & = & \ensuremath{\mathopen{(\!|}(\Varid{f}\otimes \Varid{g}) \comp \mathsf{unzip}_{\fun F}\mathclose{|\!)}}
	\label{eq:131011b}
\end{eqnarray}
where
\begin{eqnarray}
	\ensuremath{\mathsf{unzip}_{\fun F}\mathrel{=}{(\mathsf{F}\;\Varid{fst})}\kr{(\mathsf{F}\;\Varid{snd})}}
	\label{eq:131014c}
\end{eqnarray}
holds for $\ensuremath{\Varid{f}}$ and $\ensuremath{\Varid{g}}$ probabilistic and for all functors \ensuremath{\mathsf{F}} over which 
\ensuremath{\mathsf{unzip}_{\fun F}} is natural:
\begin{eqnarray}
	\ensuremath{(\mathsf{F}\;\Varid{f}\otimes \mathsf{F}\;\Varid{g}) \comp \mathsf{unzip}_{\fun F}} &=&  \ensuremath{\mathsf{unzip}_{\fun F} \comp \mathsf{F}\;(\Varid{f}\otimes \Varid{g})}	
	\label{eq:130625b}
\end{eqnarray}
\textbf{Proof:} Relying on \emph{absorption law}
\begin{eqnarray}
	(M \comp N)  \kr (P \comp Q) = (M \otimes P) \comp {(N \kr Q)} 
	\label{eq:111116c}
\end{eqnarray}
valid for any (suitably typed) matrices $M$, $N$, $P$, $Q$  \citep{Ma12ok},
we proceed by cata-universality, by solving for \ensuremath{\Varid{f}} the right hand
side equation of (\ref{eq:131014b}), once \ensuremath{\Varid{k}} is instantiated to
\ensuremath{\Varid{k}\mathrel{=}{\mathopen{(\!|}\Varid{f}\mathclose{|\!)}}\kr{\mathopen{(\!|}\Varid{g}\mathclose{|\!)}}}:
\begin{eqnarray*}
&&
	\ensuremath{({\mathopen{(\!|}\Varid{f}\mathclose{|\!)}}\kr{\mathopen{(\!|}\Varid{g}\mathclose{|\!)}}) \comp \inN }
\just={as \ensuremath{\inN } is a proper function, pair-fusion holds (\ref{eq:130129a}) }
	\ensuremath{{(\mathopen{(\!|}\Varid{f}\mathclose{|\!)} \comp \inN )}\kr{(\mathopen{(\!|}\Varid{g}\mathclose{|\!)} \comp \inN )}}
\just={two cancellations (\ref{eq:131014a}) }
	\ensuremath{{(\Varid{f} \comp \mathsf{F}\;\mathopen{(\!|}\Varid{f}\mathclose{|\!)})}\kr{(\Varid{g} \comp \mathsf{F}\;\mathopen{(\!|}\Varid{g}\mathclose{|\!)})}}
\just={pairing-absorption (\ref{eq:111116c})}
	\ensuremath{(\Varid{f}\otimes \Varid{g}) \comp ({(\mathsf{F}\;\mathopen{(\!|}\Varid{f}\mathclose{|\!)})}\kr{(\mathsf{F}\;\mathopen{(\!|}\Varid{g}\mathclose{|\!)})})}
\just={(\ref{eq:130625a}) below}
	\ensuremath{(\Varid{f}\otimes \Varid{g}) \comp \mathsf{unzip}_{\fun F} \comp \mathsf{F}\;({\mathopen{(\!|}\Varid{f}\mathclose{|\!)}}\kr{\mathopen{(\!|}\Varid{g}\mathclose{|\!)}})}
\\\ensuremath{\Box}
\end{eqnarray*}
Thus (\ref{eq:131011b}) holds, by (\ref{eq:131014b}).
As shown in the appendix, fact
\begin{eqnarray}
	\ensuremath{\mathsf{unzip}_{\fun F} \comp \mathsf{F}\;({\Varid{f}}\kr{\Varid{g}})}&= &\ensuremath{{(\mathsf{F}\;\Varid{f})}\kr{(\mathsf{F}\;\Varid{g})}}
	\label{eq:130625a}
\end{eqnarray}
used in the proof is an immediate corollary of the naturality (\ref{eq:130625b}) 
of \ensuremath{\mathsf{unzip}_{\fun F}}.
\\$\Box$
\end{theorem}

In the appendix we show that functors which support folds and for-loops are
such that (\ref{eq:130625b}) holds, thus granting ``banana-split'' (\ref{eq:131011b})
for such programming schemes. Moreover, this property is structurally preserved
by functor composition, sum etc.

In retrospect, note how law (\ref{eq:131011b}) was proved not as a corollary
of mutual recursion but as an \emph{independent} result.
Also note the major role of function \ensuremath{\mathsf{unzip}_{\fun F}} (\ref{eq:131014c}) in each inductive step:
it separates that part of the output which is to be fed to \ensuremath{\Varid{f}} from that to be fed to
\ensuremath{\Varid{g}}. It is this separation which grants \emph{non-interference} between both computations,
as happened in the \emph{square} example but not in \emph{fibonacci}
example, as we have seen.

For completeness, we state the (conditioned) mutual recursion law in a similar generic setting:

\begin{theorem}[Probabilistic mutual-recursion]\label{th:131023b}
Transformation
\begin{eqnarray}
\begin{lcbr}
	\ensuremath{\Varid{f} \comp \inN \mathrel{=}\Varid{h} \comp \mathsf{F}\;({\Varid{f}}\kr{\Varid{g}})}
\\
        \ensuremath{\Varid{g} \comp \inN \mathrel{=}\Varid{k} \comp \mathsf{F}\;({\Varid{f}}\kr{\Varid{g}})}
\end{lcbr}
	& \equiv &
	\ensuremath{{\Varid{f}}\kr{\Varid{g}}\mathrel{=}\mathopen{(\!|}{\Varid{h}}\kr{\Varid{k}}\mathclose{|\!)}}
\label{eq:130701b}
\end{eqnarray}
holds \textbf{provided} one of probabilistic $\ensuremath{\Varid{f}}$ or $\ensuremath{\Varid{g}}$ is sharp.
\\
\textbf{Proof:} generalize the rationale of section \ref{sec:130326b} from \ensuremath{\mathsf{for}}-loops to
\ensuremath{\mathsf{F}}-catamorphisms. Typically, for one such function, say \ensuremath{\Varid{f}}, to be sharp, it has to be independent
of the other (say \ensuremath{\Varid{g}}), assumed truly probabilistic. This means that \ensuremath{\Varid{h} \comp \mathsf{F}\;({\Varid{f}}\kr{\Varid{g}})\mathrel{=}\Varid{h'} \comp (\mathsf{G}\;\Varid{f})}, for some
\ensuremath{\Varid{h'}} and \ensuremath{\mathsf{G}}.
\\$\Box$
\end{theorem}

\section{Conclusions}
The production of \emph{safety critical} software is bound to
a number of certification standards in which estimating
the \emph{risk of failure} plays a central role.
NASA's procedures guide for \emph{probabilistic risk assessment} (PRA)
reviews the historical background  of risk analysis, evolving from a qualitative
to a quantitative
perspective of risk \citep{SD11}.
The UK MoD Defence Standard 00-56 \citep{MOD07} enforces that
\emph{
all (...)
calculations underpinning
the risk estimation} be recorded in so-called \emph{safety cases}
(documents supporting the claim that some given software
is safe) \emph{such that the risk estimates can be reviewed and reconstructed.}

Risk estimation 
seems to live outside programmers' core practice: either the software system
once completed is subject (by others) to intensive simulation over faults
injected into safety-critical parts, or the estimation proceeds by analysis
of worse case scenarios on a large-scale view of the system's operation.

Software development and risk analysis are performed separately because
programming language semantics are (in general) \emph{qualitative} and
risk estimation calls for \emph{quantitative} semantic models
such as those already prominent in {security} \citep{MM05}.
Quantitative methods face another problem, diagnosed by \cite{Mor12}:
probability theory is too descriptive
and not fit enough for calculation as this is understood in today's
research in program correctness.

In this paper we propose that risk calculation be constructively handled
in the programming process since the early stages, rather than being an \emph{a
posteriori} concern. This means that {risk} is taken into account
as the ``normal'' situation, absence of risk being an ideal case.
In particular, operations are modelled as probabilistic {choice}
between {expected behaviour} and faulty {behaviour}.

\emph{Functional programming} appears to be particularly apt for this purpose
because of its strong mathematical basis.
The obstacles mentioned above are circumvented by adopting a linear algebra
approach to probability calculation \citep{Ol12a}, a strategy which fits into the
calculational style of functional program development
based on its algebra of programming \citep{BM97}.

This puts functional programming in the forefront of risk estimation
simply by exploring the adjunction between distribution-valued functions and 
matrices of probabilities. One side of the adjunction is ``good for programming'':
the \emph{monadic} one, as we have shown by our experiments in Haskell; the other
side (linear algebra) is ``good for calculation''.

This does not prevent one from actually running case studies in a matrix-speaking
language such as eg.\ \matlab. Interestingly, we have observed that,
although using \matlab\ for the purposes of this paper may seem
a ``tour de force'' (since it is poorly typed and not polymorphic,
calling for explicit type error checking in the old style),
\matlab\ tends to perform faster than Haskell when
the probabilistic monadic calculations involve
distributions of wider support.

The core of this paper shows how to calculate the propagation of faults
across standard program transformation techniques known as \emph{tupling}
\citep{HITT97} and \emph{fusion} \citep{Ha11}.
This enables one to find conditions 
for the \emph{fault of the whole} to be expressed in terms of the
\emph{faults of its parts} --- a \emph{compositional} approach to risk calculation.

\section{Related and future work}

Program analysis techniques based on languages such as eg. Rely \citep{CMR13}
evaluate quantitative reliability of computations running on unreliable hardware,
eg.\ unreliable arithmetic/logical operations (as in the current paper) or
unreliable physical memories. Rely’s analysis generates \emph{reliability
pre-conditions} which are handled by \emph{reliability transformers}, bridging
to current work on probabilistic Hoare logic \citep{BGB12}.

The work by \cite{PHW10} is closer to ours in its adoption
of (untyped) linear algebra in the compositional construction of a so-called \emph{linear
operator semantics}, leading to probabilistic program
analysis inspired by classical \emph{abstract interpretation}. As in our setting,
the key element in the construction is the use of tensor products to capture different
aspects of a program.

On the foundations side, probabilistic \emph{weak} tupling has been addressed in the
more wide setting of \emph{monoidal} categories adopted in eg.\ 
categorial quantum physics \citep{Co11}. These include not only
$\mathit{FdHilb}$, the category of finite dimensional Hilbert spaces, but also
$\mathit{Rel}$, the category of binary relations. Thus the remarks
by Coecke and Paquette, in their \emph{Categories for the Practising Physicist} \citep{Co11}:
\begin{quote}\small
{$Rel$ [the category of relations] possesses more 'quantum
features' than the category $Set$ of sets and functions}
[...] 
{The categories $\mathit{FdHilb}$ and $Rel$ moreover admit a categorical matrix calculus}.
\end{quote}
We hope to exploit this connection in the future,
in particular concerning partial
orders defined for quantum states which could be used to support a notion of
refinement.

On a more practical register, we would like to find side-conditions for probabilistic
mutual-recursion (Theorem \ref{th:131023b}) weaker than that imposing
one function to be sharp. Interestingly, this seems to link to work by \cite{WB00}
on another topic: Bayesian embedded multivalued dependencies as necessary
and sufficient conditions for lossless decomposition of probabilistic relations.
For this we hope to be able to generalize previous work in this field \citep{Ol11}.  

Our experiments in probabilistic mutual recursion show that
linear versions consistently score better than the recursive.
This conforms to intuition, as program optimization leads to less
computations and therefore to lesser propagation of faults.
We would like to \emph{quantify} such a difference in probabilistic
behaviour. In general, one may think of ordering fault-injected functions with respect
to some expected, sharp function. Let \ensuremath{\Varid{f}\mathbin{:}\Conid{A}\to \Conid{B}} be such a function and \ensuremath{\Varid{g},\Varid{h}\mathbin{:}\Conid{A}\to \Conid{B}}
be probabilistic approximations to it, all represented as CS-matrices.
Then \ensuremath{\Varid{g}} and \ensuremath{\Varid{h}} can be compared against \ensuremath{\Varid{f}} as follows,
\begin{eqnarray*}
	g \leq_f h &\ensuremath{\Varid{iff}}& \ensuremath{\Varid{g} \times \Varid{f}\leq \Varid{h} \times \Varid{f}} 
\end{eqnarray*}
where \ensuremath{\Conid{M} \times \Conid{N}} denotes the Hadamard (entry-wise) product of matrices \ensuremath{\Conid{M}} and \ensuremath{\Conid{N}}.
That is, for each \ensuremath{\Varid{a}}, we compare the probability which \ensuremath{\Varid{g}} and \ensuremath{\Varid{h}} offer
for the correct value \ensuremath{\Varid{f}\;\Varid{a}}. Of course, $g \leq_f f$ always holds,
that is, \ensuremath{\Varid{f}} is the best approximation to itself.
The question is --- how effective is it to calculate with this preorder?
Is the difference \ensuremath{\Varid{h} \times \Varid{f}\mathbin{-}\Varid{g} \times \Varid{f}} a metric suitable for quantifying fault propagation
across correctness-preserving program transformations?

Scaling-up, another follow-up of the strategy put forward in this paper is its application
to fault-propagation in component-oriented software systems. \cite{CG07}
quantify component-to-component error propagation in terms of a {matrix}
which emulates a
probabilistic \emph{call-graph}. We are currently working on a formal alternative
to this approach \citep{BMO13} in which components represented by \emph{coalgebras}
\citep{lsbSBLP03} extended probabilistically, by adding to the coalgebraic
matrices of \citep{Ol12b} a \emph{behaviour} monad inside the \emph{distribution}
one.

Altogether, we hope to show that the linear algebra of programming is a wide-range
formalism suitable to generically support quantitative methods in the software sciences.

\section*{Acknowledgements}
This research was carried out in the 
QAIS (Quantitative analysis of interacting systems) 
project funded by the ERDF through the Programme COMPETE and by the Portuguese 
Government through FCT (Foundation for Science and Technology)
contract PTDC/EIA-CCO/122240/2010.

Jos\'e Oliveira wishes to thank CSW Critical Software SA
for their invitation to the final workshop of
FP7 project CriticalStep (http://www.critical-step.eu)
--- WS on Dependability and Certification ---
where the central idea of this paper was briefly presented.

Daniel Murta holds grant
{\small\text{\tt BI1\char45{}2012\char95{}PTDC\char47{}EIA\char45{}CCO\char47{}122240\char47{}2010\char95{}UMINHO}}
award-ed by FCT (Portuguese Foundation for Science and Technology).


\appendix

\section{Proofs in appendix}
\label{sec:130324a}

\paragraph{Proof of cancellation (\ref{eq:130104a})}
Base case (\ensuremath{\Varid{f}} and \ensuremath{\Varid{g}} are column vectors)
\footnote{Row vector \ensuremath{\mathbin{!}\mathbin{:}\Conid{A}\to \mathrm{1}} corresponds to the sharp,
constant function which maps every input to the singleton datatype.}:
\begin{eqnarray*}
&&
	\ensuremath{\Varid{fst} \comp ({\Varid{f}}\kr{\Varid{g}})\mathrel{=}\Varid{f}} \land \ensuremath{\Varid{snd} \comp ({\Varid{f}}\kr{\Varid{g}})\mathrel{=}\Varid{g}}
\just\equiv{\ensuremath{\Varid{fst}\mathrel{=}\Varid{id}\otimes {!}} and \ensuremath{\Varid{snd}\mathrel{=}{!}\otimes \Varid{id}}}
	\ensuremath{(\Varid{id}\otimes {!}) \comp ({\Varid{f}}\kr{\Varid{g}})\mathrel{=}\Varid{f}} \land \ensuremath{({!}\otimes \Varid{id}) \comp ({\Varid{f}}\kr{\Varid{g}})\mathrel{=}\Varid{g}}
\just\equiv{for vectors, \ensuremath{{\Varid{f}}\kr{\Varid{g}}\mathrel{=}\Varid{f}\otimes \Varid{g}} (\ref{eq:130111a}) }
	\ensuremath{(\Varid{id}\otimes {!}) \comp (\Varid{f}\otimes \Varid{g})\mathrel{=}\Varid{f}} \land \ensuremath{({!}\otimes \Varid{id}) \comp (\Varid{f}\otimes \Varid{g})\mathrel{=}\Varid{g}}
\just\equiv{functor-\ensuremath{\cdot \otimes \cdot }; natural-\ensuremath{\Varid{id}}}
	\ensuremath{\Varid{f}\otimes ({!} \comp \Varid{g})\mathrel{=}\Varid{f}} \land \ensuremath{({!} \comp \Varid{f})\otimes \Varid{g}\mathrel{=}\Varid{g}}
\just\equiv{\ensuremath{\Varid{g}} is probabilistic, therefore \ensuremath{{!} \comp \Varid{f}\mathrel{=}{!} \comp \Varid{g}\mathrel{=}{!}} \citep{Ol12a} }
	\ensuremath{\Varid{f}\otimes {!}\mathrel{=}\Varid{f}} \land \ensuremath{{!}\otimes \Varid{g}\mathrel{=}\Varid{g}}
\just\equiv{$\larrow1{\ensuremath{{!}}}1=1$ and \ensuremath{\Conid{M}\otimes \mathrm{1}\mathrel{=}\Conid{M}}}
	\ensuremath{\Varid{f}\mathrel{=}\Varid{f}} \land \ensuremath{\Varid{g}\mathrel{=}\Varid{g}}
\\\Box
\end{eqnarray*} 
Inductive step: \ensuremath{\Varid{f}\mathrel{=}[f_1|f_2]} and \ensuremath{\Varid{g}\mathrel{=}[g_1|g_2]}.
Calculating \ensuremath{\Varid{fst} \comp ({\Varid{f}}\kr{\Varid{g}})\mathrel{=}\Varid{f}} first:
\begin{eqnarray*}
&&
	\ensuremath{\Varid{fst} \comp ({\Varid{f}}\kr{\Varid{g}})\mathrel{=}\Varid{f}}
\just\equiv{\ensuremath{\Varid{f}\mathrel{=}[f_1|f_2]} and \ensuremath{\Varid{g}\mathrel{=}[g_1|g_2]}}
	\ensuremath{\Varid{fst} \comp ({[f_1|f_2]}\kr{[g_1|g_2]})\mathrel{=}[f_1|f_2]}
\just\equiv{exchange law (\ref{eq:110407a}) }
	\ensuremath{\Varid{fst} \comp [({f_1}\kr{g_1})|({f_2}\kr{g_2})]\mathrel{=}[f_1|f_2]}
\just\equiv{ fusion (\ref{eq:meither:fusion}) }
	\ensuremath{[(\Varid{fst} \comp ({f_1}\kr{g_1}))|(\Varid{fst} \comp ({f_2}\kr{g_2}))]\mathrel{=}[f_1|f_2]}
\longjust\equiv{induction hypothesis: \ensuremath{\Varid{fst} \comp ({\Varid{f}}\kr{\Varid{g}})\mathrel{=}\Varid{f}} \\ holds for \ensuremath{\Varid{f},\Varid{g}\mathbin{:=}f_i,g_i} (\ensuremath{\Varid{i}\mathrel{=}\mathrm{1},\mathrm{2}})}
	\ensuremath{[f_1|f_2]\mathrel{=}[f_1|f_2]}
\\\Box
\end{eqnarray*}
Branch \ensuremath{\Varid{snd} \comp ({\Varid{f}}\kr{\Varid{g}})\mathrel{=}\Varid{g}} is calculated in a similar way.

\paragraph{Proof of (\ref{eq:130319a})}
This equality arises from rule (\ref{eq:120428c}):
\begin{eqnarray*}
&&
	\ensuremath{\rcb\sum{\Varid{c}}{}{(\Varid{f}\;\Varid{a},\Varid{c})\;\Varid{k}\;\Varid{a}}\mathrel{=}\mathrm{1}}
\just\equiv{ one-point rule }
	\ensuremath{\rcb\sum{\Varid{b},\Varid{c}}{\Varid{f}\;\Varid{a}\mathrel{=}\Varid{b}}{(\Varid{b},\Varid{c})\;\Varid{k}\;\Varid{a}}\mathrel{=}\mathrm{1}}
\just\equiv{ \ensuremath{\Varid{b}\mathrel{=}\Varid{fst}\;(\Varid{b},\Varid{c})} ; (\ref{eq:120428c}) }
	\ensuremath{(\Varid{f}\;\Varid{a})\;(\Varid{fst} \comp \Varid{k})\;\Varid{a}\mathrel{=}\mathrm{1}}
\just\equiv{\ensuremath{\Varid{f}\mathrel{=}\Varid{fst} \comp \Varid{k}}}
	\ensuremath{(\Varid{f}\;\Varid{a})\;\Varid{f}\;\Varid{a}\mathrel{=}\mathrm{1}}
\just\equiv{\ensuremath{\Varid{f}} is sharp}
	\ensuremath{\Varid{true}}
\\ \Box
\end{eqnarray*}

\paragraph{Proof of (\ref{eq:130319c})}
This equality arises from \ensuremath{\Varid{k}} being probabilistic:
\begin{eqnarray*}
&&
	\ensuremath{\rcb\sum{\Varid{b},\Varid{c}}{\Varid{b} \not= \Varid{f}\;\Varid{a}}{(\Varid{b},\Varid{c})\;\Varid{k}\;\Varid{a}}\mathrel{=}\mathrm{0}}
\just\equiv{ \ensuremath{\mathrm{1}\mathbin{+}\mathrm{0}\mathrel{=}\mathrm{1}} }
	\ensuremath{\mathrm{1}\mathbin{+}\rcb\sum{\Varid{b},\Varid{c}}{\Varid{b} \not= \Varid{f}\;\Varid{a}}{(\Varid{b},\Varid{c})\;\Varid{k}\;\Varid{a}}\mathrel{=}\mathrm{1}}
\just\equiv{ (\ref{eq:130319a}) }
\begin{array}{l}
	\ensuremath{\rcb\sum{\Varid{c}}{}{(\Varid{f}\;\Varid{a},\Varid{c})\;\Varid{k}\;\Varid{a}}} + ~  \\ \hskip 2em
	\ensuremath{\rcb\sum{\Varid{b},\Varid{c}}{\Varid{b} \not= \Varid{f}\;\Varid{a}}{(\Varid{b},\Varid{c})\;\Varid{k}\;\Varid{a}}} = 1
\end{array}
\just\equiv{ merge quantifiers}
	\ensuremath{\rcb\sum{(\Varid{b},\Varid{c})}{}{(\Varid{b},\Varid{c})\;\Varid{k}\;\Varid{a}}} = 1
\just\equiv{ \ensuremath{\Varid{k}} is probabilistic }
	true
\\ \Box
\end{eqnarray*}
\paragraph{Proof of base-case fault propagation (\ref{eq:130220a})}
Clearly, by (\ref{eq:130325b}) and universal property (\ref{eq:130325a}), our
target (\ref{eq:130220a}) re-writes to the equality
\begin{eqnarray*}
&&
	\ensuremath{({(\mathsf{for}\;\Varid{f}\;\Varid{a})}\choice{\Varid{p}}{(\mathsf{for}\;\Varid{f}\;\Varid{b})}) \comp \inN } = \\
&& \hskip 7em
	\ensuremath{[({\underline{\Varid{a}}}\choice{\Varid{p}}{\underline{\Varid{b}}})|\Varid{f}] \comp (\mathsf{F}\;({(\mathsf{for}\;\Varid{f}\;\Varid{a})}\choice{\Varid{p}}{(\mathsf{for}\;\Varid{f}\;\Varid{b})}))}
\end{eqnarray*}
which holds by transforming the left-hand side into the right-hand side:

\begin{eqnarray*}
&&
	\ensuremath{({(\mathsf{for}\;\Varid{f}\;\Varid{a})}\choice{\Varid{p}}{(\mathsf{for}\;\Varid{f}\;\Varid{b})}) \comp \inN }
\just={ choice-fusion (\ref{eq:111216b}) }
	\ensuremath{{(\mathsf{for}\;\Varid{f}\;\Varid{a} \comp \inN )}\choice{\Varid{p}}{(\mathsf{for}\;\Varid{f}\;\Varid{b} \comp \inN )}}
\just={ (\ref{eq:130325b}) and (\ref{eq:130325a}), twice }
	\ensuremath{{([\underline{\Varid{a}}|\Varid{f}] \comp \mathsf{F}\;(\mathsf{for}\;\Varid{f}\;\Varid{a}))}\choice{\Varid{p}}{([\underline{\Varid{b}}|\Varid{f}] \comp \mathsf{F}\;(\mathsf{for}\;\Varid{f}\;\Varid{b}))})}
\just={ \ensuremath{\mathsf{F}\;\Varid{f}\mathrel{=}\Varid{id}\oplus \Varid{f}} ; $\meither M N \comp (P \oplus Q) = \meither{M\comp P}{N\comp Q}$ }
	\ensuremath{{[\underline{\Varid{a}}|(\Varid{f} \comp (\mathsf{for}\;\Varid{f}\;\Varid{a}))]}\choice{\Varid{p}}{[\underline{\Varid{b}}|(\Varid{f} \comp (\mathsf{for}\;\Varid{f}\;\Varid{b}))]}}
\just={ exchange law (\ref{eq:111219a}) }
	\ensuremath{[({\underline{\Varid{a}}}\choice{\Varid{p}}{\underline{\Varid{b}}})|({(\Varid{f} \comp \mathsf{for}\;\Varid{f}\;\Varid{a})}\choice{\Varid{p}}{(\Varid{f} \comp \mathsf{for}\;\Varid{f}\;\Varid{b})})]}
\just={ choice-fusion (\ref{eq:111219b}) }
	\ensuremath{[({\underline{\Varid{a}}}\choice{\Varid{p}}{\underline{\Varid{b}}})|(\Varid{f} \comp ({(\mathsf{for}\;\Varid{f}\;\Varid{a})}\choice{\Varid{p}}{(\mathsf{for}\;\Varid{f}\;\Varid{b})}))]}
\just={ $\meither M N \comp (P \oplus Q) = \meither{M\comp P}{N\comp Q}$ }
	\ensuremath{[({\underline{\Varid{a}}}\choice{\Varid{p}}{\underline{\Varid{b}}})|\Varid{f}] \comp (\Varid{id}\oplus ({(\mathsf{for}\;\Varid{f}\;\Varid{a})}\choice{\Varid{p}}{(\mathsf{for}\;\Varid{f}\;\Varid{b})}))}
\just={ \ensuremath{\mathsf{F}\;\Varid{f}\mathrel{=}\Varid{id}\oplus \Varid{f}} }
	\ensuremath{[({\underline{\Varid{a}}}\choice{\Varid{p}}{\underline{\Varid{b}}})|\Varid{f}] \comp (\mathsf{F}\;({(\mathsf{for}\;\Varid{f}\;\Varid{a})}\choice{\Varid{p}}{(\mathsf{for}\;\Varid{f}\;\Varid{b})}))}
\\\Box
\end{eqnarray*}

\paragraph{Proof of Khatri-Rao (conditional) fusion}
We want to prove
\begin{eqnarray}
	\ensuremath{({\Conid{M}}\kr{\Conid{N}}) \comp \Varid{h}\mathrel{=}{(\Varid{\Conid{M}.h})}\kr{(\Varid{\Conid{N}.h})}} & \implied & \mbox{\ensuremath{\Varid{h}} is sharp}
	\label{eq:130129a}
\end{eqnarray}
where probabilistic functions \ensuremath{\Varid{f}} and \ensuremath{\Varid{g}} are generalized to arbitrary matrices
\ensuremath{\Conid{M}} and \ensuremath{\Conid{N}}:
\begin{eqnarray*}
&&
	\ensuremath{(\Varid{b},\Varid{c})\;(({M}\kr{N}) \comp \Varid{h})\;\Varid{a}}
\just={ (\ref{eq:120428d}) for \ensuremath{\Varid{h}} a standard function }
	\ensuremath{(\Varid{b},\Varid{c})\;({M}\kr{N})\;(\Varid{h}\;\Varid{a})}
\just={ pointwise Khatri-Rao (\ref{eq:130129b}) }
	\ensuremath{(\Varid{b}\;M\;(\Varid{h}\;\Varid{a}))} \times \ensuremath{(\Varid{c}\;N\;(\Varid{h}\;\Varid{a}))}
\just={ (\ref{eq:120428d}) for \ensuremath{\Varid{h}} a standard function }
	\ensuremath{\Varid{b}\;(M \comp \Varid{h})\;\Varid{a}} \times \ensuremath{\Varid{c}\;(N \comp \Varid{h})\;\Varid{a}}
\just={ pointwise Khatri Rao (\ref{eq:130129b}) --- twice }
	\ensuremath{(\Varid{b},\Varid{c})\;({(M \comp \Varid{h})}\kr{(N \comp \Varid{h})})\;\Varid{a}}
\\\Box
\end{eqnarray*}

\paragraph{Proofs concerning naturality of \ensuremath{\mathsf{unzip}_{\fun F}} (\ref{eq:130625b})}
This property holds trivially for the identity functor \ensuremath{\mathsf{F}\;\Conid{X}\mathrel{=}\Conid{X}}, where
\ensuremath{\mathsf{unzip}_{\fun F}\mathrel{=}\Varid{id}}, and for any constant functor \ensuremath{\mathsf{F}\;\Conid{X}\mathrel{=}\Conid{K}}, in which case
\ensuremath{\mathsf{unzip}_{\fun F}\mathrel{=}{\Varid{id}}\kr{\Varid{id}}}.

We show next that the property is structurally preserved by functor
composition, say \ensuremath{\mathsf{F}\mathrel{=}\mathsf{G}\;\mathsf{H}}, whereby
\begin{eqnarray}
	\ensuremath{\mathsf{unzip}_{\fun{GH}}} &=& \ensuremath{\mathsf{unzip}_{\fun G} \comp (\mathsf{G}\;\mathsf{unzip}_{\fun H})}
	\label{eq:131026a}
\end{eqnarray}
holds by pair-fusion (\ref{eq:130129a}), cf.\ the sharp right term.
In this and the remaining calculations we generalize probabilistic
functions \ensuremath{\Varid{f}} and \ensuremath{\Varid{g}} in (\ref{eq:130625b}) to arbitrary matrices
\ensuremath{\Conid{M}}, \ensuremath{\Conid{N}} over a semiring. We have:
\begin{eqnarray*}
&&
	\ensuremath{\mathsf{unzip}_{\fun{GH}} \comp \mathsf{G}\;\mathsf{H}\;(\Conid{M}\otimes \Conid{N})}
\just={ (\ref{eq:131026a}) }
	\ensuremath{\mathsf{unzip}_{\fun G} \comp (\mathsf{G}\;\mathsf{unzip}_{\fun H}) \comp \mathsf{G}\;(\mathsf{H}\;(\Conid{M}\otimes \Conid{N}))}
\just={ functor \ensuremath{\mathsf{G}} (composition) }
	\ensuremath{\mathsf{unzip}_{\fun G} \comp \mathsf{G}\;(\mathsf{unzip}_{\fun H} \comp \mathsf{H}\;(\Conid{M}\otimes \Conid{N}))}
\just={ induction hypothesis: assume (\ref{eq:130625b}) for \ensuremath{\mathsf{F}\mathrel{=}\mathsf{H}}; \ensuremath{\mathsf{G}} again}
	\ensuremath{\mathsf{unzip}_{\fun G} \comp \mathsf{G}\;(((\mathsf{H}\;\Conid{M})\otimes (\mathsf{H}\;\Conid{N})) \comp (\mathsf{G}\;\mathsf{unzip}_{\fun H})}
\just={ induction hypothesis: assume (\ref{eq:130625b}) for \ensuremath{\mathsf{F}\mathrel{=}\mathsf{G}}}
	\ensuremath{((\mathsf{G}\;(\mathsf{H}\;\Conid{M}))\otimes (\mathsf{G}\;(\mathsf{H}\;\Conid{N}))) \comp \mathsf{unzip}_{\fun G} \comp \mathsf{G}\;(\mathsf{unzip}_{\fun H})}
\just={ (\ref{eq:131026a}) }
	\ensuremath{((\mathsf{G}\;\mathsf{H}\;\Conid{M})\otimes (\mathsf{G}\;\mathsf{H}\;\Conid{N})) \comp \mathsf{unzip}_{\fun{GH}}}
\\\Box
\end{eqnarray*}
Next, we do the same for sums, say \ensuremath{\mathsf{F}\mathrel{=}\mathsf{G}\oplus \mathsf{H}}. In this case we have:
\begin{eqnarray}
\ensuremath{\mathsf{unzip}_{\fun F}}&=&\ensuremath{{((\mathsf{G}\;\Varid{fst})\oplus (\mathsf{H}\;\Varid{fst}))}\kr{((\mathsf{G}\;\Varid{snd})\oplus (\mathsf{H}\;\Varid{snd}))}}
\end{eqnarray}
Facts
\begin{eqnarray}
	\ensuremath{\mathsf{unzip}_{\fun F} \comp i_1} & = & \ensuremath{(i_1\otimes i_1) \comp \mathsf{unzip}_{\fun G}}
	\label{eq:131025a}
\\
	\ensuremath{\mathsf{unzip}_{\fun F} \comp i_2} & = & \ensuremath{(i_2\otimes i_2) \comp \mathsf{unzip}_{\fun H}}
	\label{eq:131025b}
\end{eqnarray}
are easy to prove via exchange law (\ref{eq:110407a}),
where \ensuremath{i_1} and \ensuremath{i_2} are the injections of the direct sum, that is \ensuremath{[i_1|i_2]\mathrel{=}\Varid{id}}.
The same law also grants equality
\begin{eqnarray}
\begin{array}{lll}
	&& \ensuremath{[((i_1\otimes i_1) \comp ({\Conid{M}}\kr{\Conid{N}}))|((i_2\otimes i_2) \comp ({\Conid{P}}\kr{\Conid{Q}}))]} 
\\ &=& \ensuremath{{(\Conid{M}\oplus \Conid{P})}\kr{(\Conid{N}\oplus \Conid{Q})}}
\end{array}
	\label{eq:131025c}
\end{eqnarray}
which is valid for all suitably typed matrices \ensuremath{\Conid{M}}, \ensuremath{\Conid{N}}, \ensuremath{\Conid{P}} and \ensuremath{\Conid{Q}},
and will help in the proof that (\ref{eq:130625b}) holds for sums of functors
which (inductively) satisfy the same property:
\begin{eqnarray*}
&&
	\ensuremath{\mathsf{unzip}_{\fun F} \comp \mathsf{F}\;(\Conid{M}\otimes \Conid{N})}
\just\equiv{ \ensuremath{\mathsf{F}\mathrel{=}\mathsf{G}\oplus \Conid{M}} }
	\ensuremath{\mathsf{unzip}_{\fun F} \comp ((\mathsf{G}\;(\Conid{M}\otimes \Conid{N}))\oplus (\mathsf{H}\;(\Conid{M}\otimes \Conid{N})))}
\just={\ensuremath{\Conid{M}\oplus \Conid{N}\mathrel{=}[(i_1 \comp \Conid{M})|(i_2 \comp \Conid{N})]}}
	\ensuremath{[(\mathsf{unzip}_{\fun F} \comp i_1 \comp (\mathsf{G}\;(\Conid{M}\otimes \Conid{N})))|(\mathsf{unzip}_{\fun F} \comp i_2 \comp (\mathsf{H}\;(\Conid{M}\otimes \Conid{N})))]}
\just={ (\ref{eq:131025a},\ref{eq:131025b}) }
	\ensuremath{[((i_1\otimes i_1) \comp \mathsf{unzip}_{\fun G} \comp (\mathsf{G}\;(\Conid{M}\otimes \Conid{N})))|((i_2\otimes i_2) \comp \mathsf{unzip}_{\fun H} \comp (\mathsf{H}\;(\Conid{M}\otimes \Conid{N})))]}
\just={ induction hypothesis: assume (\ref{eq:130625b}) for \ensuremath{\mathsf{F}\mathrel{=}\mathsf{G}} and \ensuremath{\mathsf{F}\mathrel{=}\mathsf{H}}}
	\ensuremath{[((i_1\otimes i_1) \comp ((\mathsf{G}\;\Conid{M})\otimes (\mathsf{G}\;\Conid{N})) \comp \mathsf{unzip}_{\fun G})|((i_2\otimes i_2) \comp ((\mathsf{H}\;\Conid{M})\otimes (\mathsf{H}\;\Conid{N})) \comp \mathsf{unzip}_{\fun H})]}
\just\equiv{definitions of \ensuremath{\mathsf{unzip}_{\fun G}} and \ensuremath{\mathsf{unzip}_{\fun H}} ; absorptions}
	\ensuremath{[((i_1\otimes i_1) \comp {(\mathsf{G}\;(\Conid{M} \comp \Varid{fst}))}\kr{(\mathsf{G}\;(\Conid{N} \comp \Varid{snd}))})|((i_2\otimes i_2) \comp ((\mathsf{H}\;(\Conid{M} \comp \Varid{fst}))\otimes (\mathsf{H}\;(\Conid{N} \comp \Varid{snd}))))]}
\just={ (\ref{eq:131025c}) }
	\ensuremath{{((\mathsf{G}\;(\Conid{M} \comp \Varid{fst}))\oplus (\mathsf{H}\;(\Conid{M} \comp \Varid{fst})))}\kr{((\mathsf{G}\;(\Conid{N} \comp \Varid{snd}))\oplus (\mathsf{H}\;(\Conid{N} \comp \Varid{snd}))}}
\just\equiv{ \ensuremath{\mathsf{F}\mathrel{=}\mathsf{G}\oplus \Conid{M}} }
	\ensuremath{{(\mathsf{F}\;(\Conid{M} \comp \Varid{fst}))}\kr{(\mathsf{F}\;(\Conid{N} \comp \Varid{snd}))}}
\just\equiv{ functor \ensuremath{\mathsf{F}} ; reverse absorption }
	\ensuremath{((\mathsf{F}\;\Conid{M})\otimes (\mathsf{F}\;\Conid{N})) \comp ({(\mathsf{F}\;\Varid{fst})}\kr{(\mathsf{F}\;\Varid{snd})})}
\just\equiv{ definition of \ensuremath{\mathsf{unzip}_{\fun F}}}
	\ensuremath{((\mathsf{F}\;\Conid{M})\otimes (\mathsf{F}\;\Conid{N})) \comp \mathsf{unzip}_{\fun F}}
\\\Box
\end{eqnarray*}
Finally, we address functor \ensuremath{\mathsf{F}\;\Conid{X}\mathrel{=}\Varid{id}\otimes \Conid{X}} which participates in the recursion
schema of folds.
Note that we can replace \ensuremath{\Varid{id}} by \ensuremath{\Varid{id}\oplus \Varid{id}} since \ensuremath{\cdot \oplus \cdot } is a bifunctor
in any category of matrices \citep{Ma12ok}.
Then
\begin{eqnarray*}
	\ensuremath{\mathsf{F}\;\Conid{X}} \wider= \ensuremath{(\Varid{id}\oplus \Varid{id})\otimes \Conid{X}} \wider= \ensuremath{(\Varid{id}\otimes \Conid{X})\oplus (\Varid{id}\otimes \Conid{X})} \wider=\ensuremath{(\mathsf{F}\;\Conid{X})\oplus (\mathsf{F}\;\Conid{X})}
\end{eqnarray*}
This reduces this case to the previous one, for \ensuremath{\mathsf{G}\;\Conid{X}\mathrel{=}\mathsf{F}\;\Conid{X}} and \ensuremath{\mathsf{H}\;\Conid{X}\mathrel{=}\mathsf{F}\;\Conid{X}}, where the identities in these functors are of smaller size. Thus, in
a sense, induction proceeds on the size of the identity matrix which participates
in functor definition \ensuremath{\mathsf{F}\;\Conid{X}\mathrel{=}\Varid{id}\otimes \Conid{X}}.

\paragraph{Proof of fact (\ref{eq:130625a})}
\begin{eqnarray*}
&&
	\ensuremath{\mathsf{unzip}_{\fun F} \comp \mathsf{F}\;({\Varid{f}}\kr{\Varid{g}})}
\just={ reverse pairing-absorption (\ref{eq:111116c})}
	\ensuremath{\mathsf{unzip}_{\fun F} \comp \mathsf{F}\;(\Varid{f}\otimes \Varid{g}) \comp \mathsf{F}\;({\Varid{id}}\kr{\Varid{id}})}
\just={ naturality (\ref{eq:130625b}) }
	\ensuremath{(\mathsf{F}\;\Varid{f}\otimes \mathsf{F}\;\Varid{g}) \comp \mathsf{unzip}_{\fun F} \comp \mathsf{F}\;({\Varid{id}}\kr{\Varid{id}})}
\just={ functor \ensuremath{\mathsf{F}}; \ensuremath{\mathsf{unzip}_{\fun F}} (\ref{eq:131014c}) ; pairing-fusion (\ref{eq:130129a}), as \ensuremath{{\Varid{id}}\kr{\Varid{id}}} is sharp}
	\ensuremath{(\mathsf{F}\;\Varid{f}\otimes \mathsf{F}\;\Varid{g}) \comp ({\mathsf{F}\;(\Varid{fst} \comp ({\Varid{id}}\kr{\Varid{id}}))}\kr{\mathsf{F}\;(\Varid{fst} \comp ({\Varid{id}}\kr{\Varid{id}}))})}
\just={ standard pairing-cancellation (\ref{eq:130104a}) }
	\ensuremath{(\mathsf{F}\;\Varid{f}\otimes \mathsf{F}\;\Varid{g}) \comp ({\mathsf{F}\;\Varid{id}}\kr{\mathsf{F}\;\Varid{id}})}
\just={ functor \ensuremath{\mathsf{F}}; pairing-absorption (\ref{eq:111116c}) }
	\ensuremath{{(\mathsf{F}\;\Varid{f})}\kr{(\mathsf{F}\;\Varid{g})}}
\\\Box
\end{eqnarray*} 

\end{document}